\pdfoutput=1
\documentclass[aps,prd,twocolumn,showpacs,superscriptaddress,groupedaddress]{revtex4}  
\usepackage{graphicx}  
\usepackage{dcolumn}   
\usepackage{bm}        
\usepackage{amssymb}   
\usepackage{multirow}
\usepackage{epsfig}
\usepackage{epstopdf}
\usepackage{amsmath,booktabs,array}

\usepackage{atlasphysics}
\usepackage{eqparbox}
\usepackage{subfigure}

\newcommand{\WW}{$WW$}

\newcommand{\geant}{{\sc geant4}}

\newcommand {\sigfid}   {\ensuremath{\sigma^{\mathrm{fid}}_{WW}}}

\newcommand {\AWW}{\ensuremath{A_{WW}}}
\newcommand {\CWW}{\ensuremath{C_{WW}}}
\newcommand {\metrel}{\ensuremath{E_{\mathrm{T,~Rel}}^{\mathrm{miss}}}}
\newcommand {\metvec}{\ensuremath{\vec{E}_{\mathrm{T}}^{\mathrm{miss}}}}

\newcommand{\nWWLumi}{4.6} 
\newcommand{\nWWCrossSection}{51.9} 
\newcommand{\nWWCrossSectionStat}{2.0} 
\newcommand{\nWWCrossSectionSys}{3.9} 
\newcommand{\nWWCrossSectionLumi}{2.0} 
\newcommand{\nWWNCandEvents}{1325} 
\newcommand{\nWWNBGEvents}{369} 
\newcommand{\nWWNBGEventsUncert}{61} 
\newcommand{\nWWSMCrossSectionNLO}{44.7} 
\newcommand{\nWWSMCrossSectionUncertNLO}{$^{+2.1}_{-1.9}$} 

\usepackage{preprintcover}  
\PreprintCoverPaperTitle{Measurement of {\bf $W^+W^-$} production in $pp$ collisions at $\sqrt{s}=7$ TeV with the ATLAS detector and limits on anomalous {\bf $WWZ$} and {\bf $WW\gamma$} couplings}
\PreprintIdNumber{CERN-PH-EP-2012-242}  
\PreprintCoverAbstract{
This paper presents a measurement of the $W^+W^-$ production cross section in $pp$ collisions at $\sqrt s = 7$~TeV. 
The leptonic decay channels are analyzed using data corresponding to an integrated luminosity of \nWWLumi\ \ifb\ 
collected with the ATLAS detector at the Large Hadron Collider. The $W^+W^-$ production cross section 
$\sigma(pp\rightarrow W^+W^-+X)$ is measured to be $\nWWCrossSection \pm \nWWCrossSectionStat~{\rm (stat)}  \pm
 \nWWCrossSectionSys~{\rm (syst)}  \pm \nWWCrossSectionLumi~{\rm (lumi)}$~pb, compatible with 
the Standard Model prediction of $44.7^{+2.1}_{-1.9}$ pb.
  A measurement of the normalized fiducial cross section as a 
  function of the leading lepton transverse momentum is also presented. 
  The reconstructed transverse momentum distribution of the leading lepton is used 
  to extract limits on anomalous $WWZ$ and $WW\gamma$ couplings.}
\PreprintJournalName{Phys. Rev. D}  

\begin{document}
\title{Measurement of {\bf $W^+W^-$} production in $pp$ collisions at $\sqrt{s}=7$ TeV with the ATLAS detector and limits on anomalous {\bf $WWZ$} and {\bf $WW\gamma$} couplings}
\author{The ATLAS Collaboration}
\date{\today}

\begin{abstract}
This paper presents a measurement of the $W^+W^-$ production cross section in $pp$ collisions at $\sqrt s = 7$~TeV. 
The leptonic decay channels are analyzed using data corresponding to an integrated luminosity of \nWWLumi\ \ifb\ 
collected with the ATLAS detector at the Large Hadron Collider. The $W^+W^-$ production cross section 
$\sigma(pp\rightarrow W^+W^-+X)$ is measured to be $\nWWCrossSection \pm \nWWCrossSectionStat~{\rm (stat)}  \pm
 \nWWCrossSectionSys~{\rm (syst)}  \pm \nWWCrossSectionLumi~{\rm (lumi)}$~pb, compatible with 
the Standard Model prediction of  \nWWSMCrossSectionNLO\ \nWWSMCrossSectionUncertNLO~pb.
  A measurement of the normalized fiducial cross section as a 
  function of the leading lepton transverse momentum is also presented. 
  The reconstructed transverse momentum distribution of the leading lepton is used 
  to extract limits on anomalous $WWZ$ and $WW\gamma$ couplings. 
\end{abstract}

\maketitle

\section{Introduction}
Measurements of vector boson pair production at particle colliders provide 
important tests of the electroweak sector of the Standard Model (SM). Deviations of the 
production cross section or of kinematic distributions from their SM predictions could arise from 
anomalous triple gauge boson interactions~\cite{theo1} or from new particles decaying into vector bosons~\cite{theo2}.
Vector boson pair production at the Large Hadron Collider (LHC)~\cite{lhc} also represents an important 
source of background to Higgs boson production~\cite{HWW} and to searches for physics beyond the SM. 

\begin{figure}[h]
 \centering
 \subfigure[]{\includegraphics[width=0.2\textwidth]{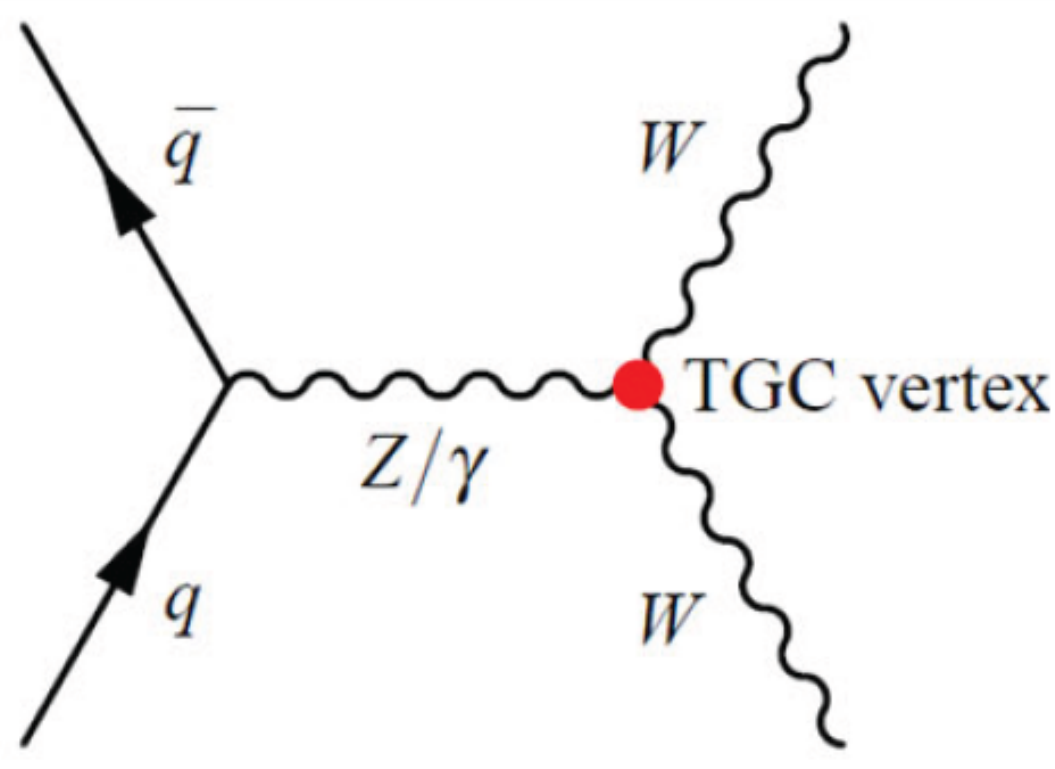}} \hspace{1cm}
 \subfigure[]{\includegraphics[width=0.15\textwidth]{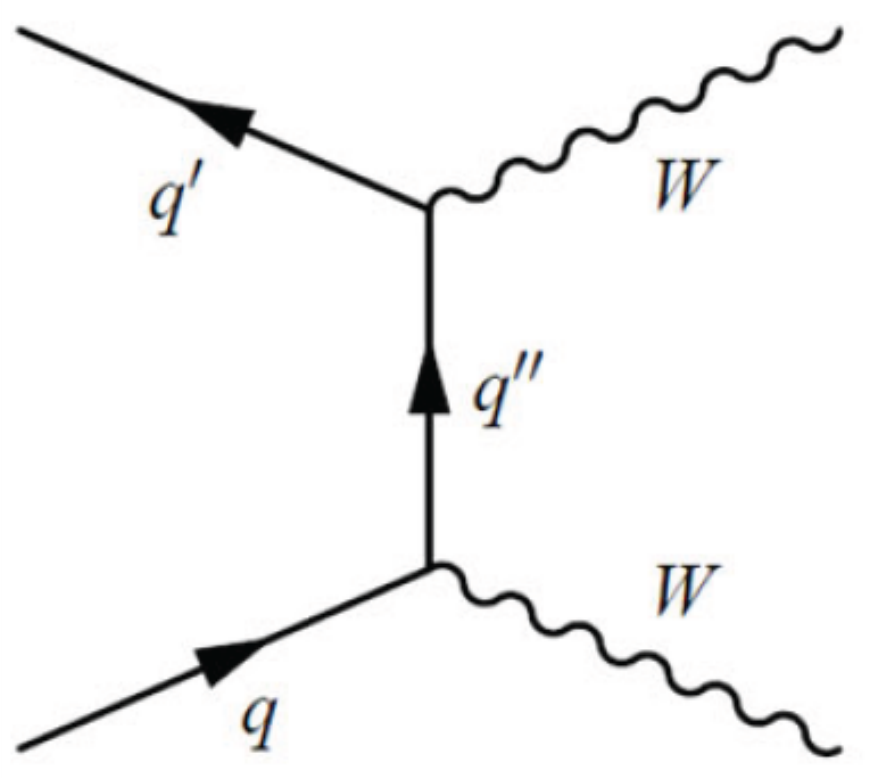}} 
 \caption{SM LO Feynman diagrams for \WW\ production 
 through the $q\bar q$ initial state at the LHC for (a) the $s$-channel and (b) the $t$-channel. The
 $s$-channel diagram contains the $WWZ$ and $WW\gamma$ TGC vertices. } 
 \label{fig:qq2ww_prod}
\end{figure}

This paper describes a measurement of the $W^+W^-$ (hereafter \WW) inclusive and differential production cross sections and 
limits on anomalous $WWZ$ and $WW\gamma$ triple gauge couplings (TGCs) in purely leptonic decay 
channels $WW \rightarrow \ell\nu\ell'\nu'$ with $\ell, \ell'=e, \mu$.
$WW \rightarrow \tau\nu \ell\nu$ and $WW \rightarrow \tau\nu\tau\nu$ processes with 
$\tau$ leptons decaying into electrons or muons with additional neutrinos are also included. 
Three final states are considered based on the lepton flavor, namely $ee$, $\mu\mu$, and $e\mu$.  
Leading-order (LO) Feynman diagrams for \WW\ production at the LHC 
include $s$-channel production with either a $Z$ boson or a virtual photon as the mediating particle or $u$- and 
$t$-channel quark exchange. 
The $s$- and $t$-channel diagrams are shown in Fig.~\ref{fig:qq2ww_prod}. 
Gluon-gluon fusion processes involving box diagrams contribute about 3\% to the total cross section. 
The SM cross section for $WW$ production in $pp$ collisions at $\sqrt{s}=7$~TeV is predicted at next-to-leading order (NLO) to be 
\nWWSMCrossSectionNLO \nWWSMCrossSectionUncertNLO~pb. 
The calculation of the total cross section is performed 
using {\sc mcfm}~\cite{mcfm} with the {\sc ct10}~\cite{cteq} parton distribution functions (PDFs).
An uncertainty of $^{+4.8\%}_{-4.2\%}$ is evaluated based on the variation of renormalization ($\mu_R$) and factorization ($\mu_F$) scales 
by a factor of two ($^{+3.6\%}_{-2.5\%}$) and {\sc ct10} PDF uncertainties derived from the eigenvector error 
sets as described in Ref.~\cite{pdfuncert} ($^{+3.1\%}_{-3.4\%}$) added in quadrature. 
The contribution from SM Higgs production~\cite{HWW} with the Higgs boson decaying into a pair of $W$ bosons 
($H \rightarrow WW$) depends on the mass of the Higgs boson ($m_H$). For $m_H=126$ GeV, the 
SM $WW$ production cross section would be increased by 8\%. 
Contributions from vector boson fusion (VBF) and double parton scattering (DPS)~\cite{dps} processes are found 
to be less than 0.1\%. The processes involving the SM Higgs boson, VBF and DPS are not 
included neither in the \WW\ cross-section predictions, nor in deriving the corrected measured cross sections. 
Events containing two $W$ bosons from top-quark pair production and single top-quark production 
are explicitly excluded from the signal definition, 
and are treated as background contributions. 

The $s$-channel diagram contains the $WWZ$ and $WW\gamma$ couplings. 
The SM predicts that these couplings are $g_{WWZ}=-e \cot \theta_W$ and $g_{WW\gamma}=-e$, where 
$e$ is related to the fine-structure constant $\alpha$ ($=e^2/4\pi$) and $\theta_W$ is the weak mixing angle.
Detailed studies of \WW\ production allow to test the non-Abelian structure of the SM electroweak theory and 
probe anomalous $WWZ$ and $WW\gamma$ TGCs, which may be sensitive to low-energy manifestations of new 
physics at a higher mass scale.
\WW\ production and anomalous $WWZ$ and $WW\gamma$ TGCs have been previously 
studied by the LEP~\cite{lep} and Tevatron~\cite{tev} experiments, 
and were also recently studied by the LHC experiments~\cite{atlasww1, atlasww2, cmsww}.  
The dataset used in this paper corresponds to an integrated luminosity of \nWWLumi~\ifb~\cite{lumi} collected 
with the ATLAS detector at the LHC, and the results presented supersede the previous ATLAS measurements~\cite{atlasww2}.

This paper is organized as follows. Section II describes the overall analysis strategy. 
Section III describes the ATLAS detector. Section IV summarizes the Monte Carlo (MC) simulation 
used for the signal and background modeling. Section V details the reconstruction of final state 
objects and event selection criteria. Sections VI and VII describe the $WW$ signal and background 
estimation. Results are presented in Sec. VIII for inclusive and fiducial cross sections; 
in Sec. IX for the normalized differential fiducial cross section as a function of the transverse 
momentum ($p_{\rm T}$)~\cite{coordinate} of the lepton with higher $p_{\rm T}$ (denoted by the ``leading lepton"); 
and in Sec. X for limits on anomalous $WWZ$ and $WW\gamma$ TGCs. Conclusions are drawn in Sec. XI. 

\section{Analysis Strategy}
Candidate \WW\ events are selected with two opposite-sign charged leptons (electrons or muons) and 
large missing transverse momentum (\met), a signature referred to ``$\ell\ell' +\met$" in this paper. 
The cross section is measured in a fiducial phase space and also in the total phase space. 
The fiducial phase space is defined in Sec. VI and is chosen to be close to the phase space defined 
by the offline selection criteria. 
The fiducial cross section $\sigma^{\rm fid}_{WW}$ for the $pp \rightarrow WW  + X \rightarrow \ell \nu \ell' \nu' + X$ 
process is calculated according to the equation
\begin{equation}
   \sigma_{WW}^{\rm fid} = \frac{N_{\rm data} - N_{\rm bkg}}{\CWW \times {\cal{L}}}\ ,
\label{eqn:fidXsection}
\end{equation}

\noindent where $N_{\rm data}$ and $N_{\rm bkg}$ are the number of observed data events and estimated background events, respectively. 
\CWW\ is defined as the ratio of the number of events satisfying all offline selection criteria to the number of 
events produced in the fiducial phase space and is estimated from simulation. $\cal{L}$ is the integrated luminosity of the data sample. 

The total cross section $\sigma_{WW}$ for the $pp \rightarrow WW + X$ process is calculated for each channel using the equation
\begin{equation}
   \sigma_{WW} = \frac{N_{\rm data} - N_{\rm bkg}}{\CWW \times A_{WW} \times {\rm BR} \times {\cal{L}}}\ ,
\label{eqn:totalXsection}
\end{equation}
\noindent where \AWW\ represents the kinematic and geometric acceptance from the total phase space to the fiducial phase space, 
and BR is the branching ratio for both $W$ bosons decaying into $e\nu$ or $\mu\nu$ (including decays through $\tau$ leptons with additional neutrinos). 
The combined total cross section from the three channels is determined by minimizing a negative log-likelihood 
function as described in Sec. VIII. 

To obtain the normalized differential $WW$ cross section in the fiducial phase space ($1/\sigfid \times d \sigfid/dp_{\rm T}$),
the reconstructed leading lepton $p_{\rm T}$ distribution is corrected for detector effects after the subtraction of 
background contamination. The measured leading lepton $p_{\rm T}$ spectrum is also used to extract anomalous $WWZ$ and $WW\gamma$ TGCs. 

\section{The ATLAS Detector}
The ATLAS detector~\cite{atlas} is a multi-purpose particle physics detector with approximately forward-backward symmetric cylindrical 
geometry. The inner detector (ID) system is immersed in a 2~T axial magnetic field and provides tracking information 
for charged particles in the pseudorapidity range $|\eta|<2.5$. 
It consists of a silicon pixel detector, a silicon microstrip detector, and a transition radiation tracker. 

The calorimeter system covers the pseudorapidity range $|\eta|<4.9$. The highly segmented electromagnetic calorimeter 
consists of lead absorbers with liquid-argon (LAr) as active material and covers the pseudorapidity range $|\eta|<3.2$. 
In the region $|\eta|<1.8$, a pre-sampler detector using a thin layer of LAr is used to correct for the 
energy lost by electrons and photons upstream of the calorimeter. 
The electron energy resolution is about $2-4$\% at $\pt=45$ GeV.
The hadronic tile calorimeter is a steel/scintillating-tile 
detector and is situated directly outside the envelope of the electromagnetic calorimeter. 
The two endcap hadronic calorimeters have LAr as the active material and copper absorbers. The calorimeter coverage is 
extended to $|\eta|=4.9$ by a forward calorimeter with LAr as active material and copper and tungsten as absorber material.
The jet energy resolution is about 15\% at $\pt=45$ GeV.

The muon spectrometer measures the deflection of muons in the large superconducting air-core toroid magnets. It covers 
the pseudorapidity range $|\eta|<2.7$ and is instrumented with separate trigger and high-precision tracking chambers. 
A precision measurement of the track coordinates in the principal bending direction of the magnetic field is provided 
by drift tubes in the pseudorapidity range $|\eta|<2.0$. At large pseudorapidities, cathode strip chambers 
with higher granularity are used in the innermost plane over $2.0<|\eta|<2.7$. The muon trigger system, which covers 
the pseudorapidity range $|\eta|<2.4$, consists of resistive plate chambers in the barrel ($|\eta|<1.05$) 
and thin gap chambers in the endcap regions ($1.05<|\eta|<2.4$). 
The muon momentum resolution is about $2-3$\% at $\pt=45$ GeV.

A three-level trigger system is used to select events for offline analysis. 
The level-1 trigger is implemented in hardware and uses a subset of detector 
information to reduce the event rate to a design value of at most 75 kHz. This is followed by two
software-based trigger levels, level-2 and the event filter, which together reduce the event rate 
to about 400~Hz which is recorded for analysis.

\section{Monte Carlo Simulation}
\label{sec:mc} 
Signal \WW\ events are modeled using MC-simulated samples, while 
contributions from various SM background physics processes are estimated using a combination of MC samples and control samples from data. 
MC events are generated at $\sqrt{s} = 7$~TeV and processed through the full detector simulation~\cite{atlassim} based on \geant~\cite{geant}. 
The simulation includes the modeling of additional $pp$ interactions in the same and neighboring bunch crossings. 

The simulation of the \WW\ signal production is based on samples of $q\bar{q} \rightarrow WW$ and $gg \rightarrow WW$ 
events generated with {\sc mc@nlo}~\cite{mcatnlo} and {\sc gg2ww}~\cite{gg2ww}, respectively. Initial parton momenta are 
modeled with the {\sc ct10} PDFs. 
The parton showering and hadronization, and the underlying event are modeled with {\sc herwig}~\cite{herwig} and {\sc jimmy}~\cite{jimmy}.

The SM background processes, which are described in Sec. VII, are simulated using 
{\sc alpgen}~\cite{alpgen} for the $W$+jets, Drell-Yan $Z/\gamma^*$+jets and 
$W\gamma$ processes, {\sc mc@nlo} for the $t\bar{t}$ process, {\sc madgraph}~\cite{madgraph} for 
the $W\gamma^*$ process, {\sc acermc}~\cite{acermc} for the single top-quark process, and {\sc herwig} 
for $WZ$ and $ZZ$ processes. 
The {\sc tauola}~\cite{tauola} and {\sc photos}~\cite{photos} programs are used to model the decay 
of $\tau$ leptons and QED final-state radiation of photons, respectively.  
The MC predictions are normalized to the data sample based on the integrated luminosity and 
cross sections of the physics processes. Higher-order corrections, if available, are applied.
The cross section is calculated to next-to-next-to-leading-order (NNLO) accuracy for $W$ 
and $Z/\gamma^*$~\cite{fewz}, NLO plus next-to-next-to-leading-log order for $t\bar{t}$~\cite{ttbar}, and NLO for 
$WZ$ and $ZZ$ processes~\cite{mcfm}. 

To improve the agreement between data and simulation, lepton selection efficiencies are measured
in both data and simulation, and correction factors are applied to the simulation to account for differences with respect to data. 
Furthermore, the simulation is tuned to reproduce the calorimeter energy and the muon momentum scale and resolution observed in data. 

\section{Objects and Event Selection}
\label{sec:event_selections} 
The data analyzed were selected online by a single-lepton 
($e$ or $\mu$) trigger with a threshold on the transverse energy  in the electron 
case and on the transverse momentum in the muon case. 
Different thresholds (18~GeV for muons and 20~GeV or 22~GeV for electrons) were applied for different running periods. 
After applying data quality requirements, the total integrated luminosity  
is \nWWLumi~fb$^{-1}$ with an uncertainty of 3.9\% for all three channels $ee$, $\mu\mu$ and $e\mu$~\cite{lumi}. 

Due to the presence of multiple $pp$ collisions in a single bunch crossing, each event can 
have multiple vertices reconstructed. The primary vertex of the hard collision is defined as the vertex with the highest 
$\sum p_{\rm T}^{2}$ of associated ID tracks. 
To reduce contamination due to cosmic rays, the primary vertex must have 
at least three associated tracks with $\pt > 0.4$~GeV. 
 
Electrons are reconstructed from a combination of an electromagnetic cluster in the calorimeter and a track in the ID, 
and are required to have $\pt > 20$~GeV and lie within the range $|\eta| < 2.47$, excluding the transition region 
between the barrel and endcap calorimeters ($1.37 < |\eta| < 1.52$). 
The electron \pt\ is calculated using the energy measured in the electromagnetic calorimeter and the track direction 
measured by the ID. 
Candidate electrons must satisfy the tight quality definition~\cite{eid} re-optimized for 2011 data-taking conditions,  
which is based on the calorimeter shower shape, 
track quality, and track matching with the calorimeter cluster. 

Muon candidates must be reconstructed in the ID and the muon spectrometer, and
the combined track is required to have $\pt > 20$~GeV and $|\eta|<2.4$. 
Good quality reconstruction is ensured by requiring minimum numbers of silicon microstrip and
pixel hits associated with the track~\cite{muonreco}. 

To ensure candidate electrons and muons originate from the primary 
interaction vertex, they are also required to have a longitudinal impact parameter ($|z_0|$) smaller than 1 mm and a
transverse impact parameter ($|d_0|$) divided by its resolution ($\sigma_{d_0}$) smaller than ten for electrons and three for muons. 
These requirements reduce contamination from heavy-flavor quark decays and cosmic rays. 

To suppress the contribution from hadronic jets which are misidentified as leptons, 
electron and muon candidates are required to be isolated in both the ID and the calorimeter. 
The sum of transverse energies of all clusters around the lepton but not associated with the lepton 
within a cone of size $\Delta R = \sqrt{(\Delta\eta)^2 + (\Delta\phi)^2}= 0.3$ is required to be less than 14\% of the lepton transverse momentum. 
Corrections to the sum of transverse energies of all clusters around the lepton are applied to account for the energy deposition 
inside the isolation cone due to electron energy leakage or muon energy deposition and additional $pp$ collisions.
The sum of the \pt\ of all tracks with $\pt > 1$ GeV that originate from the primary vertex and are 
within a cone of size $\Delta R = 0.3$ around the lepton track is required to be less than 13\% (15\%) of the electron (muon) \pt.
 
Jets are reconstructed from topological clusters of energy in the calorimeter using the anti-$k_t$ 
algorithm~\cite{antikt} with radius parameter $R = 0.4$. 
Topological clustering extends up to $|\eta|=4.9$, and clusters are seeded 
by calorimeter cell deposits having energy exceeding 4 standard deviations of the cell noise level. 
Jet energies are calibrated using $\pt$- and $\eta$-dependent correction factors based on the simulation, 
and validated by collision data studies~\cite{jet_study}. 
Jets are classified as originating from $b$-quarks by using an algorithm that combines information 
about the impact parameter significance of tracks in a jet which has a topology of semileptonic $b$- 
or $c$-hadron decays~\cite{bjet}. The efficiency of the $b$-tagging algorithm is 85\% for $b$-jets 
in $t\bar{t}$ events, with an average light jet rejection factor of 10.

Since electrons are also reconstructed as jets, if a reconstructed jet and an electron satisfying 
the criteria mentioned above lie within $\Delta R=0.3$ of each other, the jet is discarded. 
Electrons and muons are required to be separated from each other by $\Delta R>0.1$. 
Since muons can radiate photons which can convert to electron-positron pairs, if a muon and an electron lie within $\Delta R=0.1$ of 
each other, the electron is discarded.

The measurement of the missing transverse momentum two-dimensional vector $\vec{E}_{\rm T}^{\rm miss}$ and 
its magnitude \met\ is based on the measurement of the energy collected by the electromagnetic and hadronic calorimeters, 
and muon tracks reconstructed by the ID and the muon spectrometer. Calorimeter cells associated with reconstructed jets 
with $\pt\ > 20$~GeV are calibrated at the hadronic energy scale, whereas calorimeter cells not associated with any object 
are calibrated at the electromagnetic energy scale.

Events with exactly two oppositely-charged leptons passing the lepton selection criteria above are selected.
At least one of the two leptons is required to be geometrically matched to a lepton reconstructed by the trigger algorithm. 
In order to ensure that the lepton trigger efficiency reaches its plateau region and does not depend on 
the $p_{\rm T}$ of the lepton, the matching lepton is required to have $p_{\rm T}> 25$~GeV. 
The leading lepton $p_{\rm T}$ requirement also helps to reduce the $W$+jets background contribution.  

Events satisfying the above requirements are dominated by the contribution from the Drell-Yan process.
To reject this background contribution, different requirements on the dilepton invariant mass $m_{\ell \ell'}$ 
and a modified missing transverse energy, \metrel, are applied to each final state. 
The \metrel\ variable is defined as:
\begin{equation}
\label{eqn:MetRelDef}
\metrel = \left\{ 
\begin{array} {ll}
\met \times \sin\left(\Delta\phi\right) & \mbox{if} \; \Delta\phi < \pi/2 \\
\met                                              & \mbox{if} \; \Delta\phi \geq \pi/2 
\end{array}
\right.
\end{equation}

\noindent where $\Delta\phi$ is the difference in the azimuthal angle between the $\vec{E}_{\rm T}^{\rm miss}$ and the 
nearest lepton or jet.
The \metrel\ variable is designed to reject events where the apparent \met\ arises from a mismeasurement 
of lepton momentum or jet energy.  
The selection criteria applied to $m_{\ell \ell'}$ and \metrel\ are $m_{\ell \ell'}>15, 15, 10$~GeV, 
$|m_{\ell \ell'}-m_Z|>15, 15, 0$~GeV, and $\metrel > 45, 45, 25$~GeV for the $ee$, $\mu\mu$ and $e\mu$ channels, respectively. 
Less strict selection criteria on $m_{\ell\ell'}$ and \metrel\ are employed for the $e\mu$ channel since the contribution from the 
Drell-Yan process is inherently smaller. 

With the application of the $m_{\ell \ell'}$ and \metrel\ selection criteria, the remaining background events 
come mainly from $t\bar{t}$ and single top-quark processes. To reject this background contribution,  
events are vetoed if there is at least one jet candidate with $p_{\rm T}>25$~GeV and $|\eta|<4.5$ 
(this selection criterion is denoted by the term ``jet veto'' in this paper). 
To further reduce the Drell-Yan contribution, the transverse momentum of the dilepton system, 
$p_{\rm T}(\ell\ell')$, is required to be greater than 30~GeV for all three channels. 

Figures~\ref{fig:mll_compare}--\ref{fig:ptll_compare} show comparisons between data and simulation for the 
$m_{\ell\ell'}$, \metrel, jet multiplicity and $p_{\rm T}(\ell\ell')$ distributions before the successive cuts are applied 
to the $ee$, $\mu\mu$ and $e\mu$ channels, respectively.
The contributions from various physics processes are estimated using MC simulation and
normalized to the cross sections as described in Sect. IV.
These plots indicate the discrimination power of these variables to reduce the dominant 
$t\bar{t}$, $W+$jets and Drell-Yan backgrounds and improve the signal-to-background ratio. Discrepancies between data and SM predictions based on pure MC estimates for some plots indicate the need for data-driven background estimates as are used for the $WW$ signal extraction.

\begin{figure*}
  \begin{center}
    \subfigure[]{\includegraphics[width=0.33\textwidth]{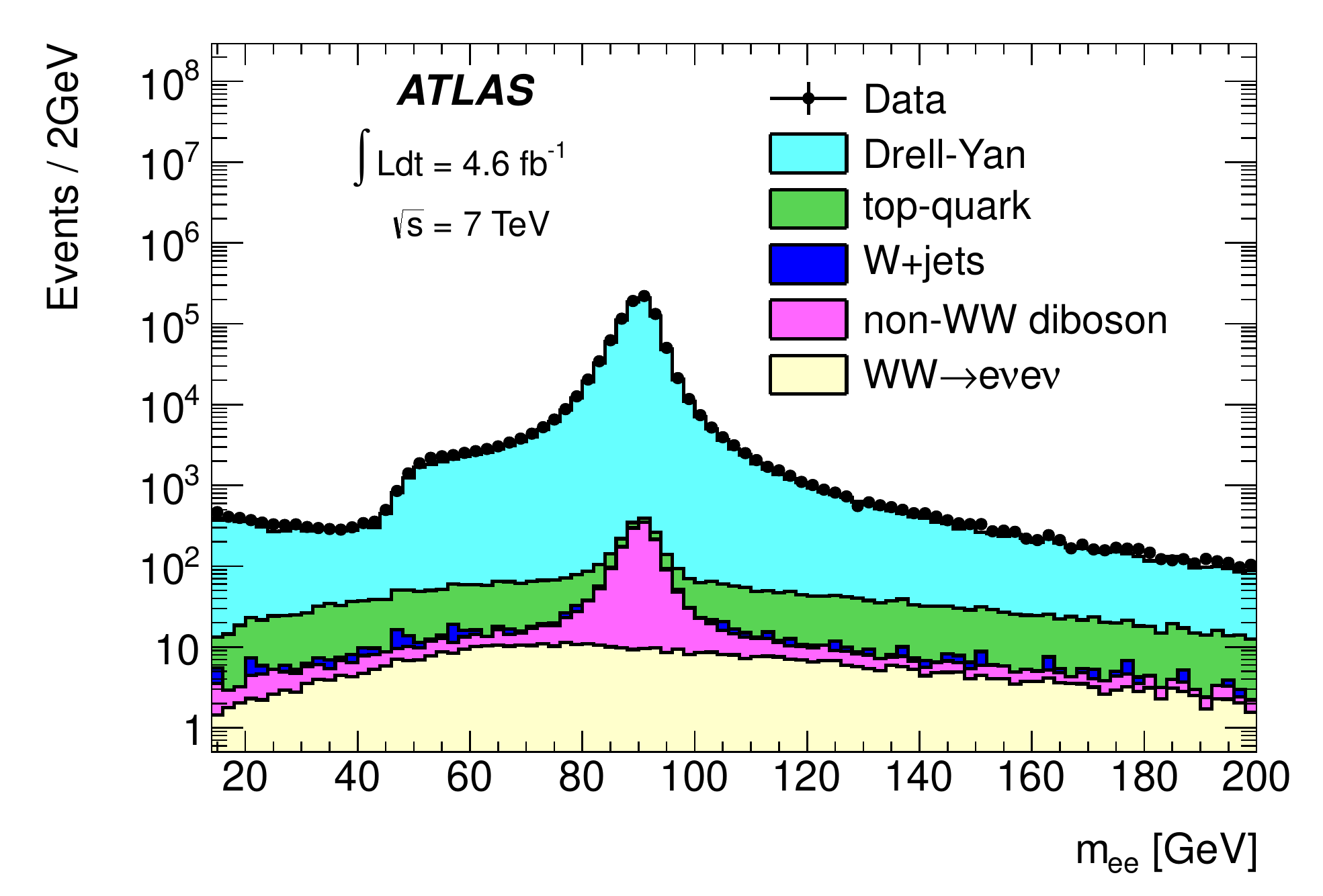}}  
    \subfigure[]{\includegraphics[width=0.33\textwidth]{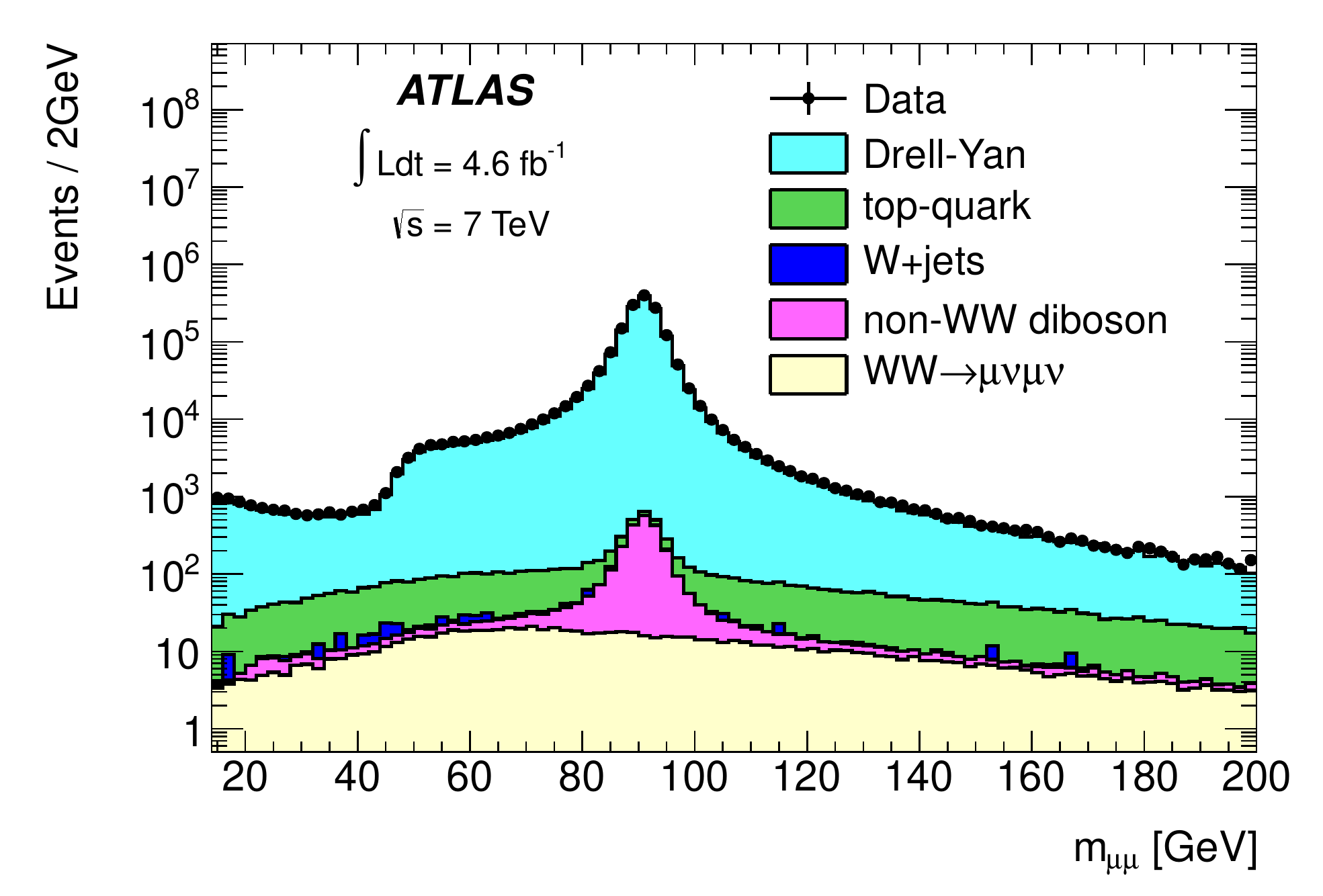}}
    \subfigure[]{\includegraphics[width=0.33\textwidth]{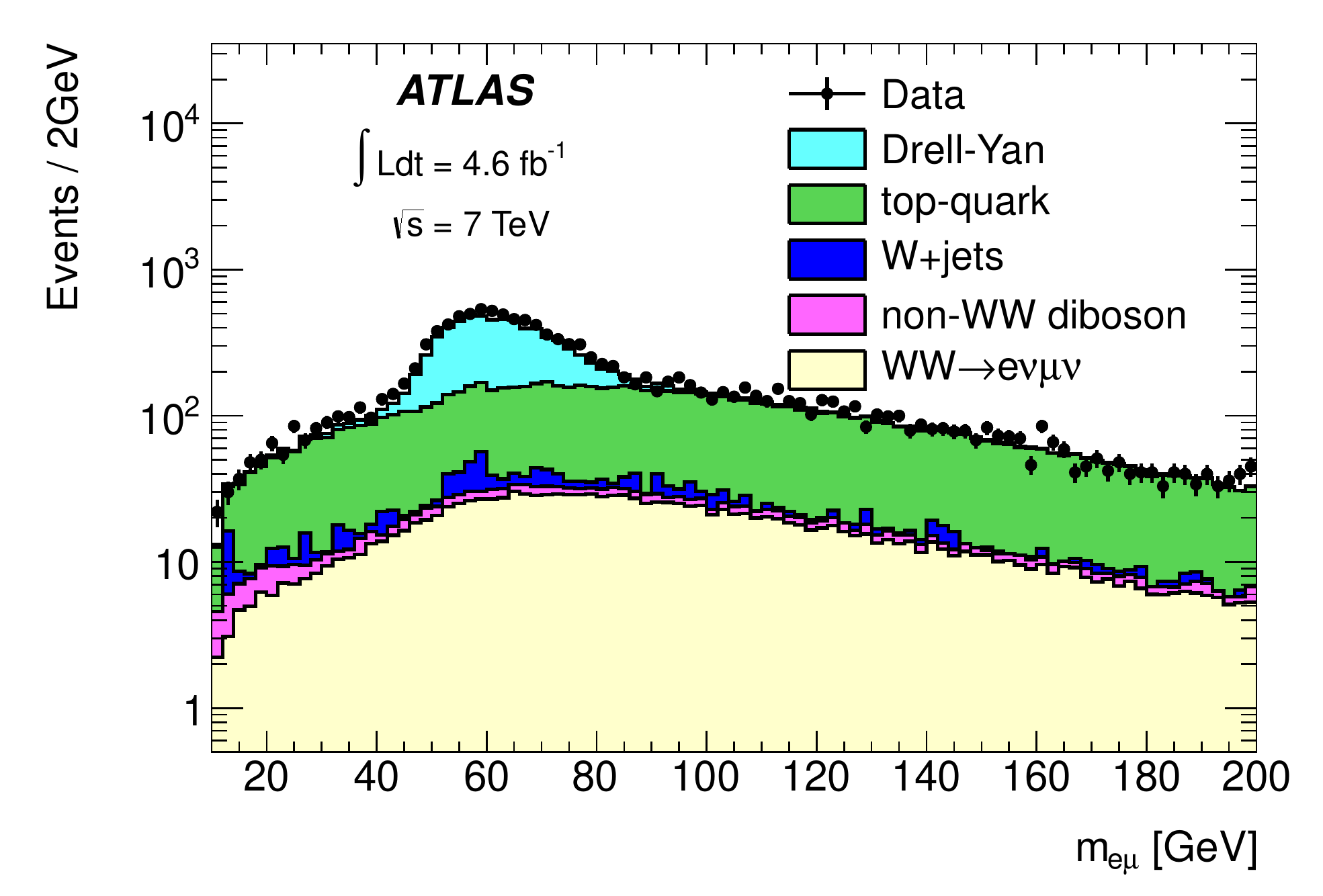}} 
   \caption{Comparison between data and simulation for the dilepton invariant mass distribution before the $m_{\ell\ell'}$ cut for the (a) $ee$, (b) $\mu\mu$ and (c) $e\mu$ channels, respectively. The contributions from various physics processes are estimated using MC simulation and normalized to the cross sections as described in Sect. IV.}
 \label{fig:mll_compare}
 \end{center}
 \end{figure*}

\begin{figure*}
  \begin{center}
    \subfigure[]{\includegraphics[width=0.33\textwidth]{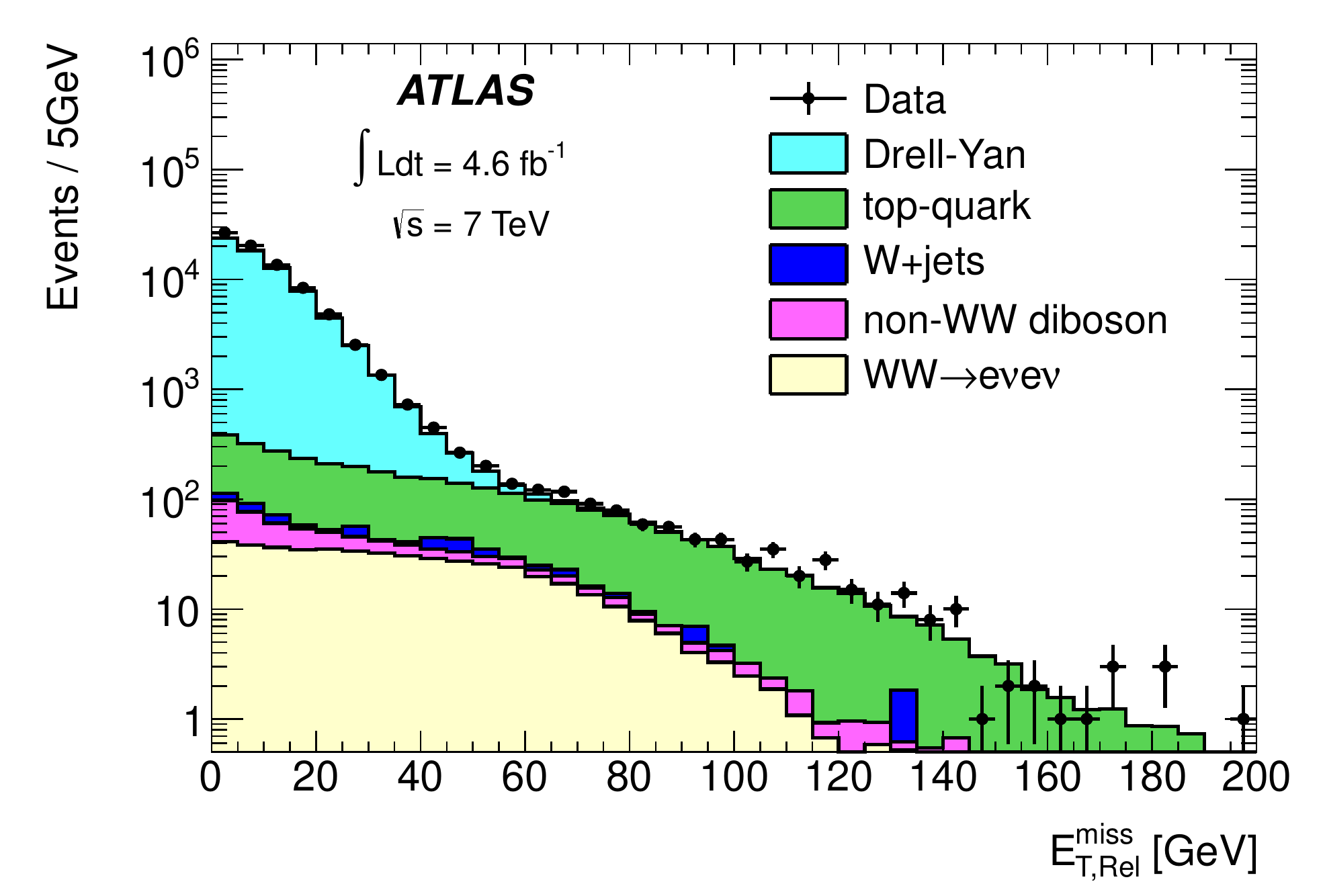}}  
    \subfigure[]{\includegraphics[width=0.33\textwidth]{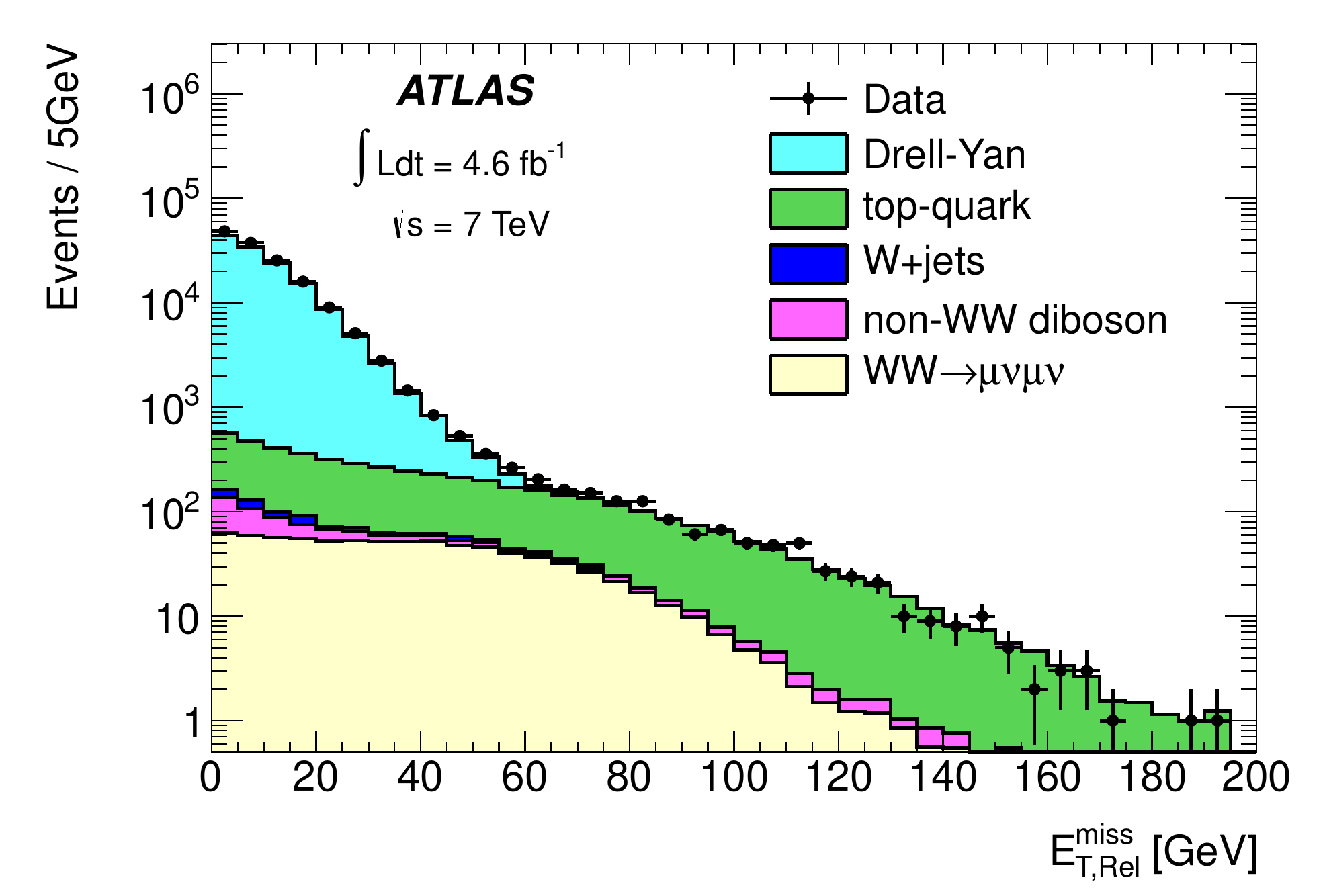}}
    \subfigure[]{\includegraphics[width=0.33\textwidth]{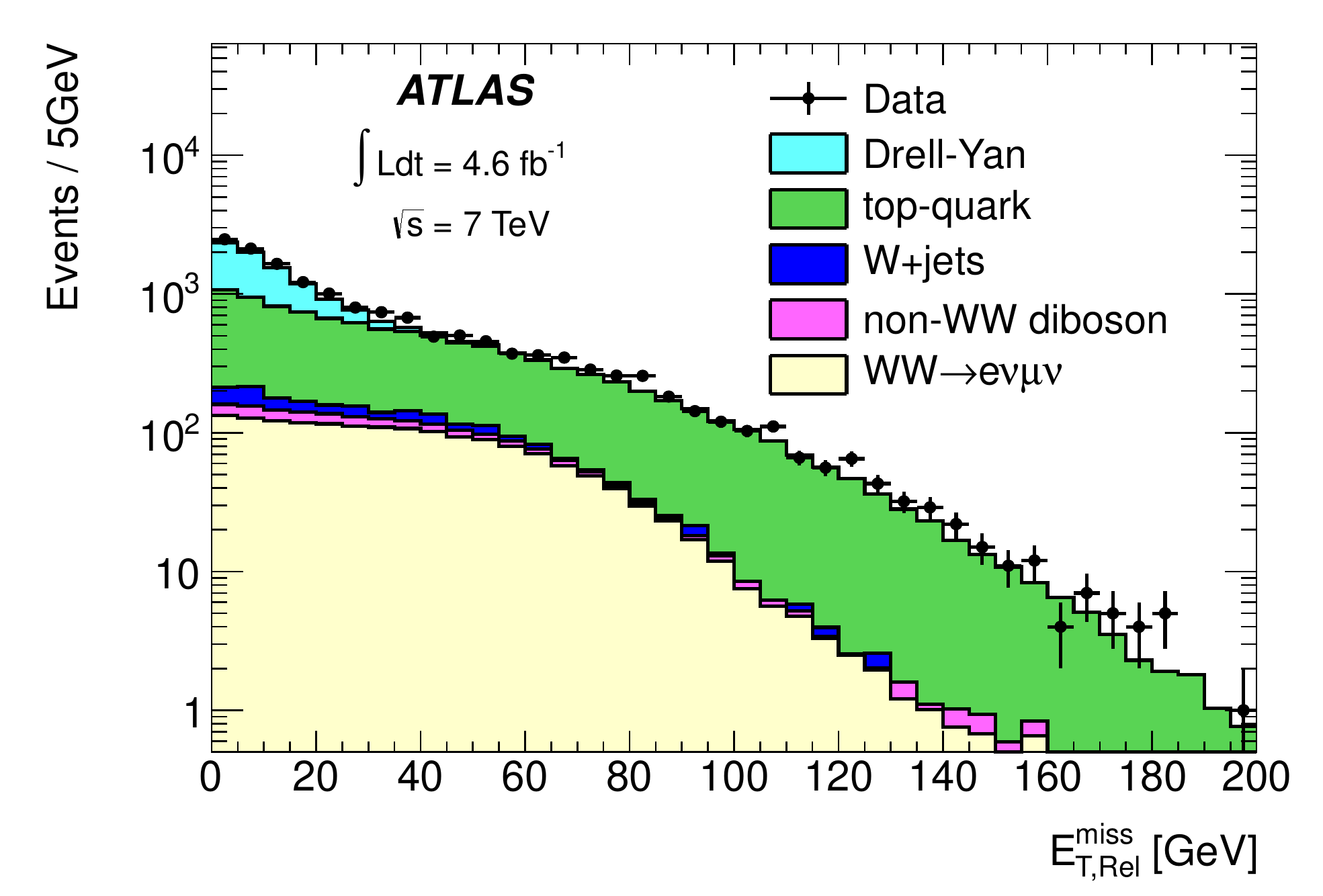}}   
    \caption{Comparison between data and simulation for the \metrel\ distribution before the \metrel\ cut for the (a) $ee$, (b) $\mu\mu$ and (c) $e\mu$ channels, respectively. The contributions from various physics processes are estimated using MC simulation and normalized to the cross sections as described in Sect. IV.}
 \label{fig:metrel_compare}
 \end{center}
 \end{figure*}

\begin{figure*}
  \begin{center}
    \subfigure[]{\includegraphics[width=0.33\textwidth]{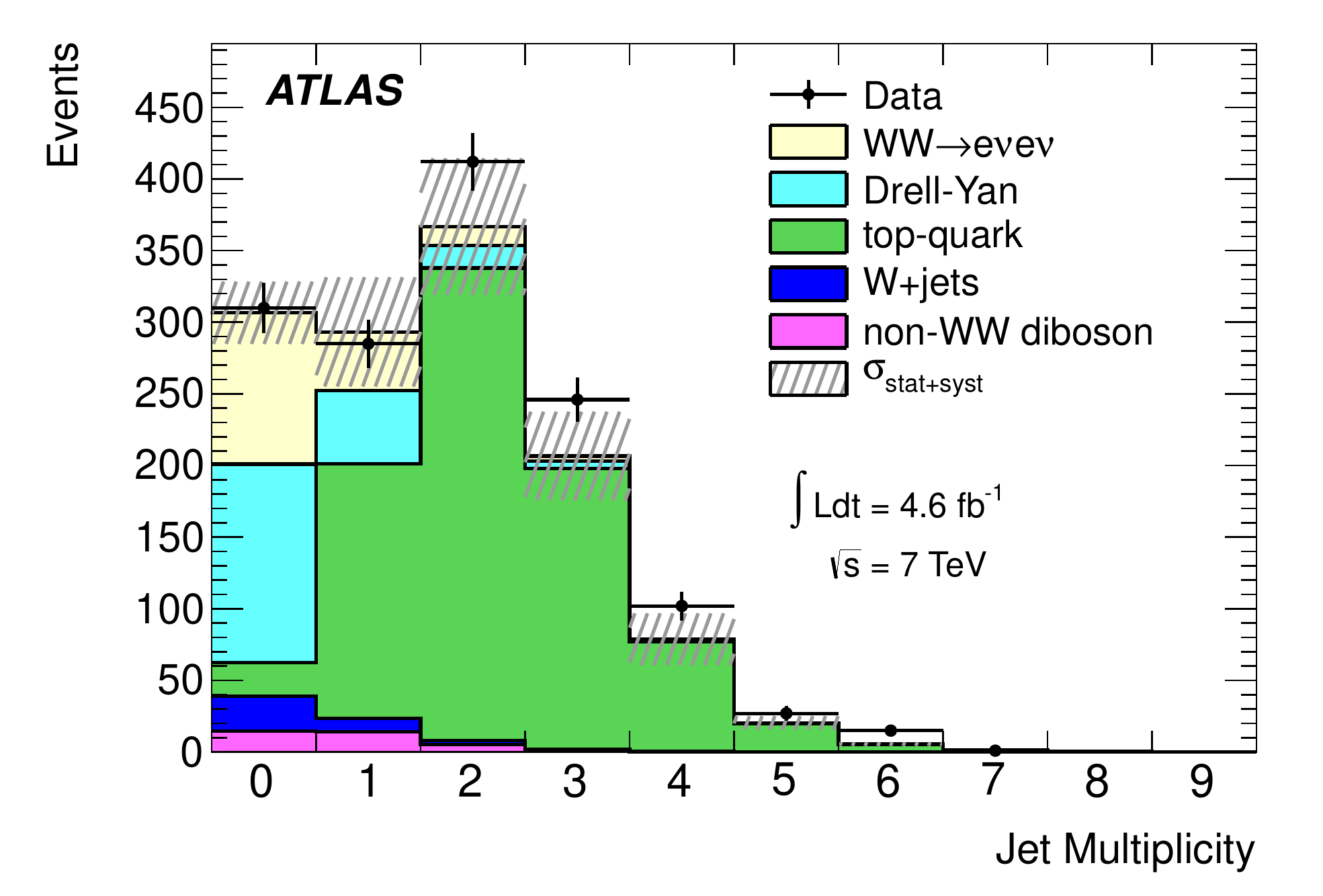}}  
    \subfigure[]{\includegraphics[width=0.33\textwidth]{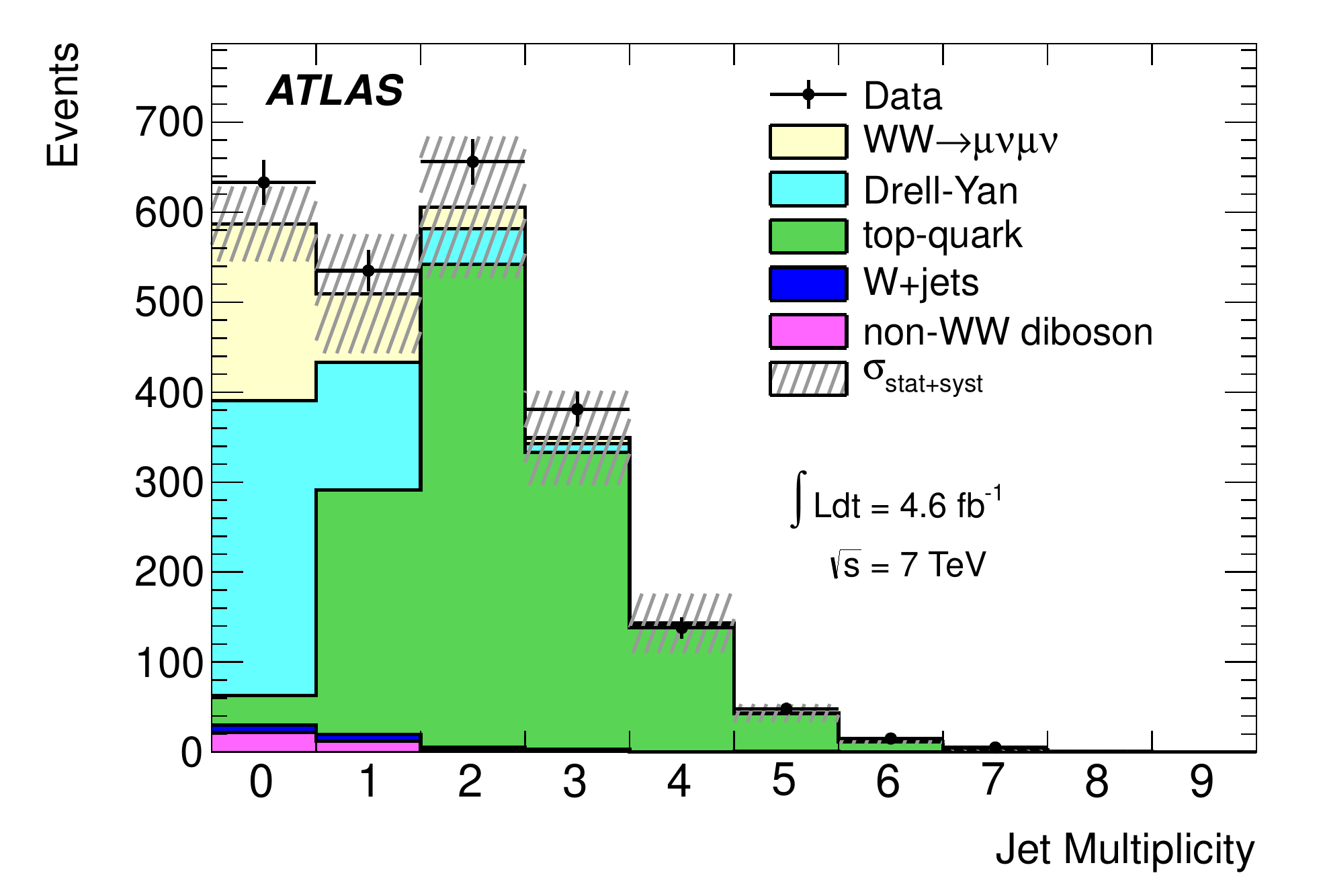}}
    \subfigure[]{\includegraphics[width=0.33\textwidth]{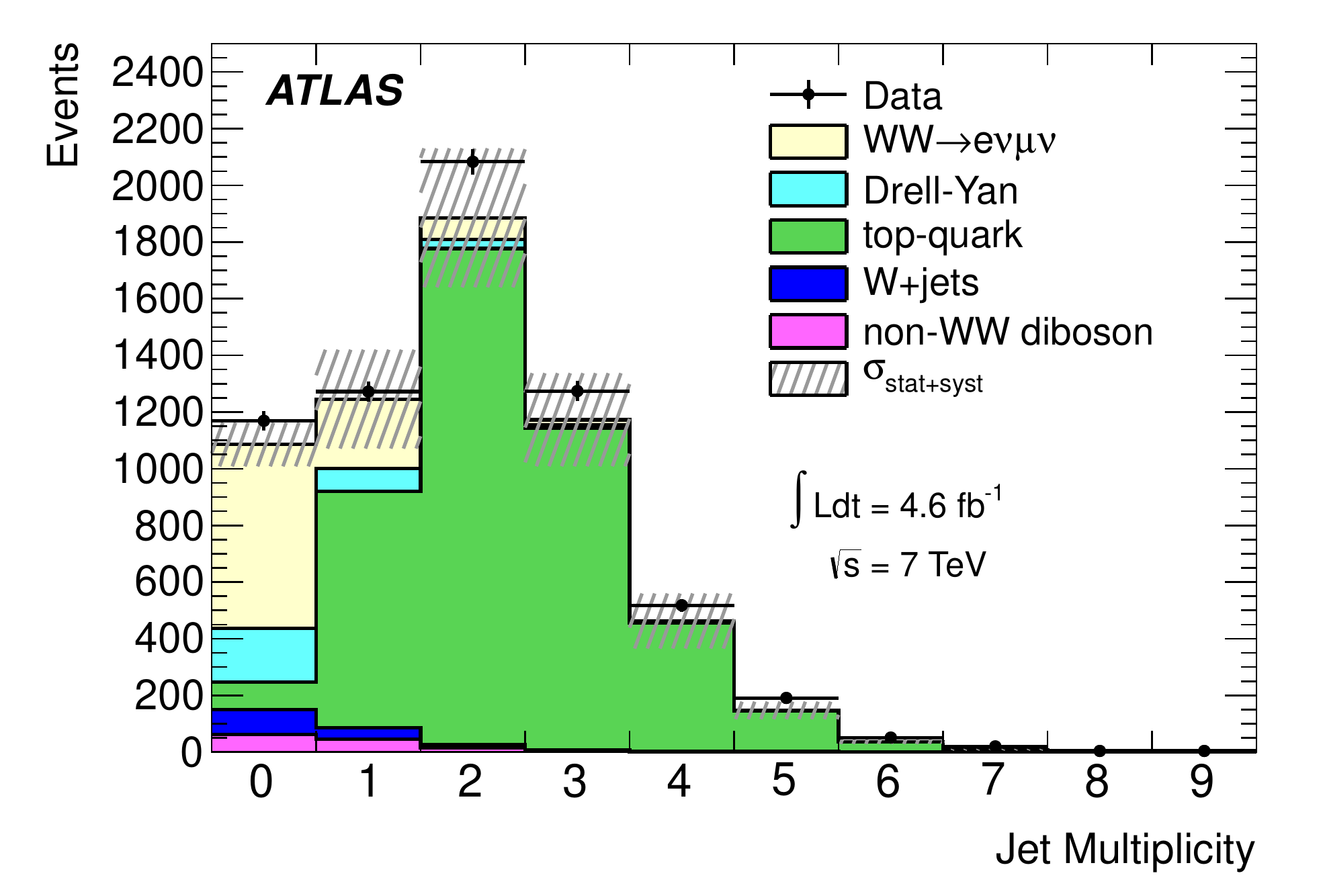}} 
    \caption{Comparison between data and simulation for the jet multiplicity distribution of jets with $p_{\rm T}>25$ GeV before jet veto requirement for the (a) $ee$, (b) $\mu\mu$ and (c) $e\mu$ channels, respectively. The contributions from various physics processes are estimated using MC simulation and normalized to the cross sections as described in Sect. IV. 
The error band on each plot includes both statistical and systematic uncertainties on the signal and background estimations. Systematic uncertainties on the signal estimation are described in Sect. VI. 
Systematic uncertainties on background estimations include uncertainties on lepton, jet and \met\ reconstruction and identification, as well as uncertainties on theoretical production cross sections for these processes.}
\label{fig:njets_compare}
 \end{center}
 \end{figure*}

\begin{figure*}
  \begin{center}
    \subfigure[]{\includegraphics[width=0.33\textwidth]{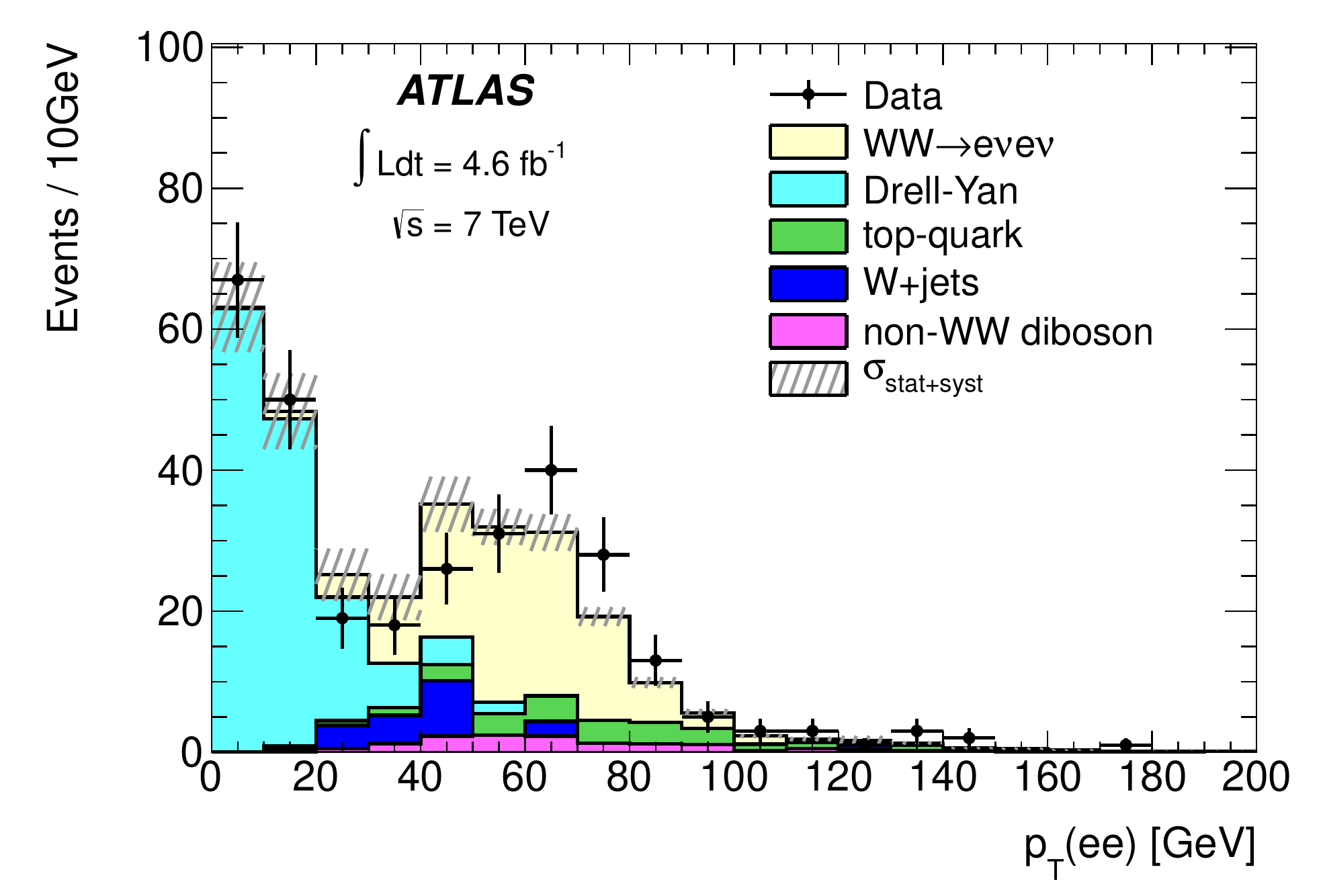}}  
    \subfigure[]{\includegraphics[width=0.33\textwidth]{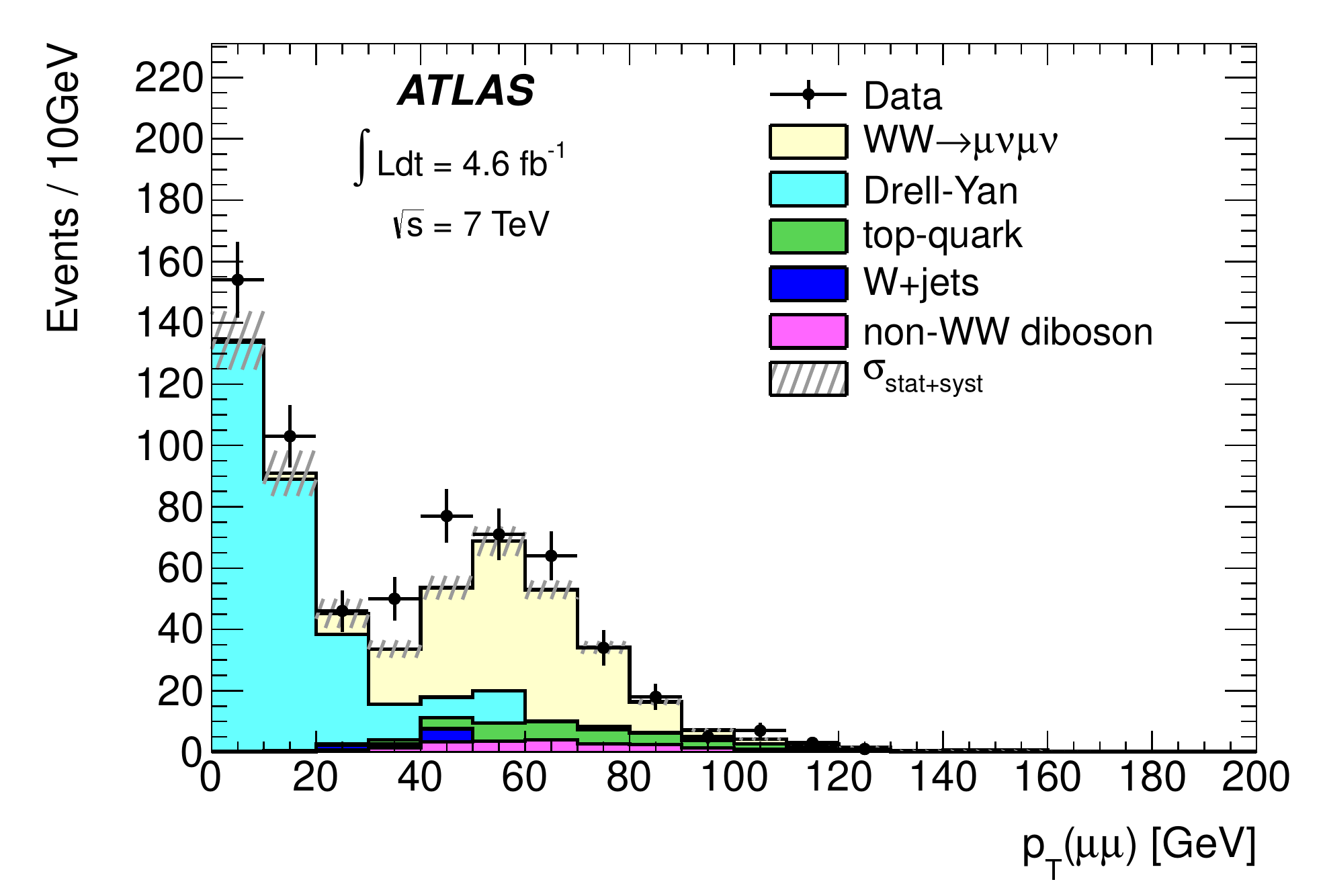}}
    \subfigure[]{\includegraphics[width=0.33\textwidth]{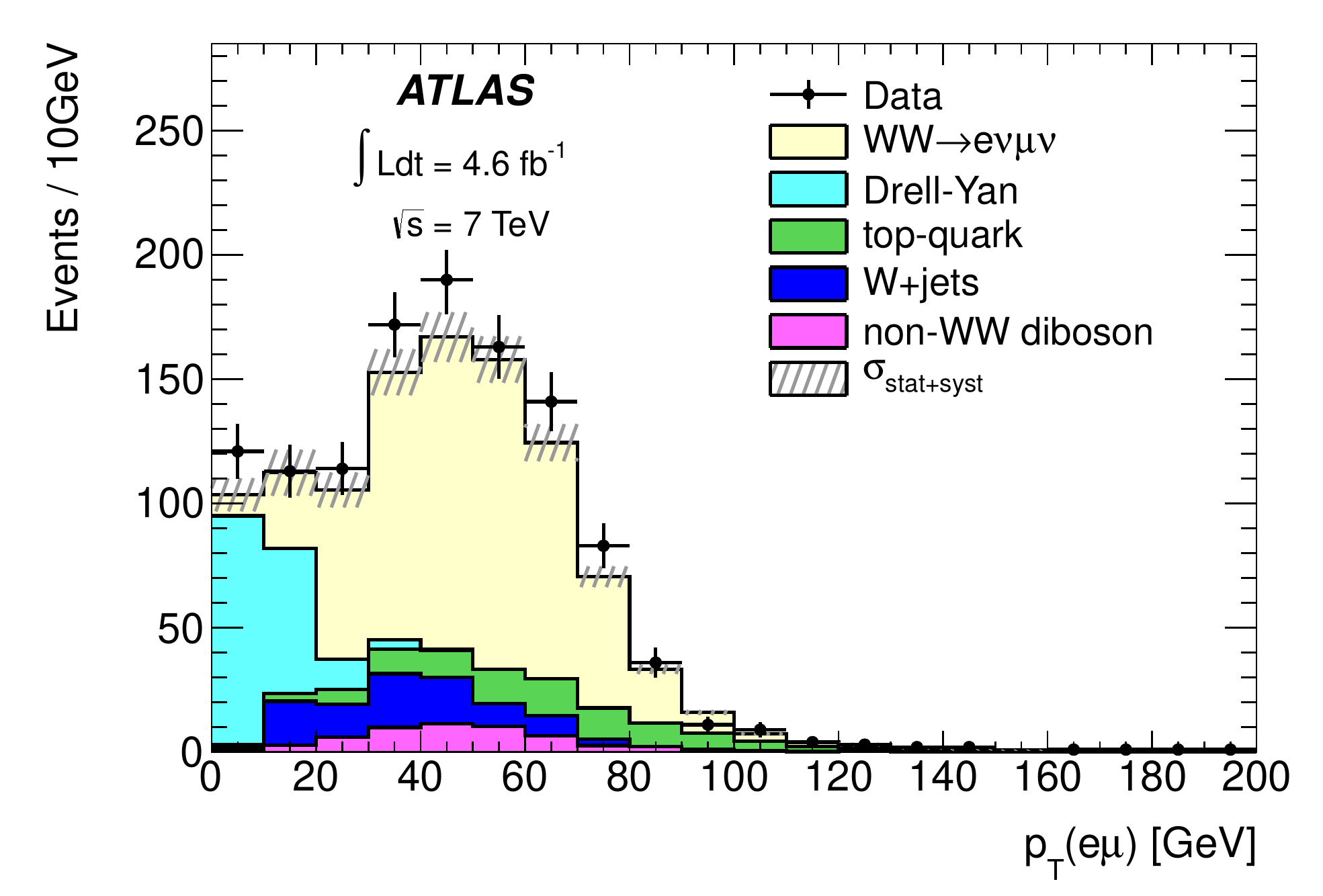}} 
    \caption{Comparison between data and simulation for the dilepton $p_{\rm T}$ distribution before the $p_{\rm T}(\ell\ell')$ cut for the (a) $ee$, (b) $\mu\mu$ and (c) $e\mu$ channels, respectively. The contributions from various physics processes are estimated using MC simulation and normalized to the cross sections as described in Sect. IV. 
The error band on each plot includes both statistical and systematic uncertainties on the signal and background estimations. Systematic uncertainties on the signal estimation are described in Sect. VI. 
Systematic uncertainties on background estimations include uncertainties on lepton, jet and \met\ reconstruction and identification, as well as uncertainties on theoretical production cross sections for these processes.}
 \label{fig:ptll_compare}
 \end{center}
 \end{figure*}

\section{$WW$ Signal Acceptance}
The fractions of simulated \WW\ signal events remaining after each step of the event selection are summarized in Table~\ref{ta:cutflow_mc_acc}. 
The fractions for direct $WW$ decays into electrons or muons are shown separately from processes involving $\tau$ leptons 
($WW \rightarrow \tau\nu \ell\nu$ and $WW \rightarrow \tau\nu \tau\nu$ processes with $\tau$ leptons decaying into electrons or muons). 
The acceptance for the $\mu\mu$ channel is higher than the $ee$ channel since the identification efficiency 
for muons is higher than that for electrons. The acceptance for the $e\mu$ channel is the highest one due to looser 
selection requirements applied to $m_{\ell \ell'}$ and \metrel. 

In order to minimize the theoretical uncertainty due to the extrapolation from the 
measured phase space to the total phase space for the cross-section measurement, 
a fiducial phase space is defined at the generator level by selection criteria similar to those used offline. 
Generator-level jets are reconstructed by running the anti-$k_t$ algorithm with radius parameter $R=0.4$ 
on all final-state particles generated with the {\sc mc@nlo} and {\sc gg2ww} event generators after parton showering and hadronization.  
The fiducial phase space is defined with the following criteria: lepton $p_{\rm T}>20$~GeV, 
muon pseudorapidity $|\eta|<2.4$, electron pseudorapidity $|\eta|<1.37$ or $1.52<|\eta|<2.47$, 
no generator-level jets with $p_{\rm T}>25$~GeV, 
rapidity $|y|<4.5$ and separated from an electron by $\Delta R>0.3$. 
The leading lepton $p_{\rm T}$ is required to be above 25~GeV and $p_{\rm T}^{\ell \ell'}>30$~GeV. 
The events are further required to have $m_{\ell \ell'}>15, 15, 10$~GeV, $|m_{\ell \ell'}-m_Z|>15, 15, 0$~GeV, 
and $p_{\rm T,~Rel}^{\nu+\bar{\nu}} > 45, 45, 25$~GeV for the $ee$, $\mu\mu$ and $e\mu$ channels respectively. 
The $p_{\rm T,~Rel}^{\nu+\bar{\nu}}$ variable is defined similarly to \metrel, where the $\vec{E}_{\rm T}^{\rm miss}$ is replaced by the vector sum 
of the $\vec{p}_{\rm T}$ of the two generator-level neutrinos. 
To reduce the dependence on QED radiation, the electron and muon $p_{\rm T}$ include 
contributions from photons within $\Delta R=0.1$ of the lepton direction.

With this definition of the fiducial phase space, the overall acceptance times efficiency can be separated into two factors $A_{WW}$ and 
$C_{WW}$, where $A_{WW}$ represents the extrapolation from the fiducial phase space to the total phase space, while 
$C_{WW}$ represents detector effects such as lepton trigger and identification efficiencies, with a small contribution from differences in 
generated and measured phase spaces due to detector resolution. 

Corrections to the simulation of lepton identication efficiencies 
and resolutions are discussed in Sec.~\ref{sec:mc}. 
A correction to the modelling of the jet veto efficiency (the fraction of events with 
zero reconstructed jets) is determined as the ratio of data to MC jet veto efficiencies for the $Z/\gamma^* \rightarrow \ell\ell$ process.  
This ratio is applied to $WW$ MC~\cite{campbell} as 
\begin{equation}
P^{WW}_{\rm pred}= \frac{P_{Z/\gamma^*}^{\rm data}}{P_{Z/\gamma^*}^{\rm MC}} \times P_{WW}^{\rm MC}, 
\end{equation}
where $P^{WW}_{\rm pred}$ is the corrected jet veto efficiency for $pp \rightarrow WW$, $P_{WW}^{\rm MC}$ is the 
MC estimate of this efficiency, and $P^{\rm data}_{Z/\gamma^*}$ ($P^{\rm MC}_{Z/\gamma^*}$) is the efficiency 
determined using $Z/\gamma^* \rightarrow \ell \ell$ events selected with two leptons satisfying the lepton 
selection criteria and $|m_{\ell\ell}-m_Z|<15$~GeV in data (MC).  
By applying this correction, experimental uncertainties 
associated with the jet veto efficiency are significantly reduced, in particular, the 
uncertainty on the jet energy scale.  The dominant uncertainty is 
due to the theoretical prediction of the differences in jet energy 
spectra between the $WW$ and $Z/\gamma^*$ processes, which are both modelled with 
{\sc mc@nlo} $+$ {\sc herwig} for this correction.

For the factor $C_{WW}$ ($A_{WW}$), the dominant 
uncertainty is the theoretical uncertainty on 
$P^{\rm MC}_{Z/\gamma^*}$ ($P^{\rm MC}_{WW}$). The theoretical uncertainty from missing higher-order corrections is 
evaluated by varying renormalization and factorization scales up and down by a factor of 2 for both the inclusive 
($\ge 0$) and exclusive ($\ge 1$) jet cross sections and adding these two uncertainties in 
quadrature~\cite{Stewart_Tackmann}. Uncertainties associated with the parton shower and hadronization models 
are evaluated by comparing the {\sc pythia} and {\sc herwig} models, interfaced to the MC generating the process 
of interest.  Uncertainties due to PDFs are computed using the {\sc ct10} error eigenvectors, and using the 
difference between the central {\sc ct10} and {\sc mstw2008nlo}~\cite{mstw} PDF sets.  Including 
uncertainties from the jet energy scale (JES) and jet energy resolution (JER), $P^{WW}_{\rm pred}$ is estimated 
to be $0.624 \pm 0.023$, $0.625 \pm 0.023$ and $0.633 \pm 0.023$ for the $ee$, $\mu\mu$ and $e\mu$ channels, 
respectively.  

Additional theoretical uncertainties on $A_{WW}$ are evaluated using the same procedures as for the jet veto 
efficiency.  Additional uncertainties on $C_{WW}$ are calculated using uncertainties on the lepton trigger, 
reconstruction and isolation efficiencies, as well as energy scale and resolution uncertainties on the reconstruction 
of lepton, jet, soft clustered energy in the calorimeter, and energy deposits from additional $pp$ collisions. 
The uncertainty on the single-lepton trigger efficiency is less than 0.5\%~\cite{trig_uncertainty}. Electron 
and muon reconstruction and identification efficiency uncertainties are less than 2.0\% and 0.4\%, 
respectively~\cite{eff_uncertainty}.  The lepton isolation efficiency is determined with an uncertainty of 
0.3\% and 0.2\% for electrons and muons, respectively.  The simulation is corrected for the differences with 
respect to the data in lepton energy scale and resolution.  The uncertainty is less than 1.0\% and 0.1\% on 
the energy scale and less than 0.6\% and 5.0\% on the resolution, for electrons and muons, respectively~\cite{eid}.  
Uncertainties on the JES range from 2.5\% to 8\%, varying with jet $p_{\rm T}$ and $\eta$~\cite{jet_uncertainty}.  
Uncertainties on the JER range from 9--17\% for jet $p_{\rm T} \simeq 30$~GeV to about 5--9\% for jets with 
$p_{\rm T}>180$~GeV depending on jet $\eta$~\cite{jet_uncertainty}.  The uncertainties on the lepton energy scale 
and resolution, JES and JER are propagated to the \met, which also receives contributions from energy deposits 
due to additional $pp$ collisions in the same or close by bunch crossings, and from energy deposits not associated 
with any reconstructed object~\cite{dmet}. 

All systematic uncertainties described above are propagated to the calculations of $A_{WW}$, $C_{WW}$ and $A_{WW} \times C_{WW}$. 
The overall systematic uncertainty on $A_{WW}$ is 5.7\% for all three channels.  
The contributions from all systematic sources for $A_{WW}$ are listed in Table~\ref{tbl:aww2}.   
The overall systematic uncertainty on $C_{WW}$ is 4.2\%, 3.1\% and 3.2\% for the $ee$, $\mu\mu$ and $e\mu$ channels, respectively. 
The contributions from all systematic sources for $C_{WW}$ are listed in Table~\ref{tbl:cww}. 

The product of $A_{WW} \times C_{WW}$ is defined as the ratio of events satisfying all offline selection criteria 
to the number of events produced in the total phase space. The systematic uncertainty on $A_{WW} \times C_{WW}$ 
is 4.9\%, 4.0\% and 4.1\% for the $ee$, $\mu\mu$ and $e\mu$ channels. 
Owing to the presence of correlations between $A$ and $C$, these uncertainties are smaller than those 
obtained by adding in quadrature the uncertainties from the PDFs, $\mu_F$, $\mu_R$, and parton shower model. As a result, the uncertainty on $A_{WW} \times C_{WW}$ is used for the calculation 
of the total cross-section uncertainty in each individual channel.  
Table~\ref{tbl:awwcww} summarizes the central value and also the statistical and systematic uncertainties on $A_{WW}$, $C_{WW}$ and $A_{WW} \times C_{WW}$ 
for all three channels. 

\begin{table*}
  \centering
  \begin{tabular}{lrr|rr|rr} \hline \hline
    Selection criteria & \multicolumn{2}{c|}{$ee$} & \multicolumn{2}{c|}{$\mu\mu$} & \multicolumn{2}{c}{$e\mu$} \\ \cline{2-7}
                              & $e\nu e\nu$  & $\tau\nu \ell\nu$& $\mu\nu \mu\nu$ & $\tau\nu \ell\nu$& $e\nu \mu\nu$ & $\tau\nu \ell\nu$ \\ \hline
    Exactly two opposite-sign leptons~~~~~  & 22.8\%  &  7.3\%  & 39.0\% & 11.4\%  &  30.2\% & 9.1\% \\ 
    $m_{\ell\ell'}>15, 15, 10$ GeV   & 22.7\%  &  7.3\%  & 38.8\% & 11.4\%  &  30.2\% & 9.1\% \\
    $|m_{\ell\ell'}-m_Z|>15, 15, 0$ GeV      & 17.6\%  &  5.4\%   & 29.9\% & 8.5\%  &  30.2\% & 9.1\% \\
    $\metrel>45, 45, 25$ GeV                 & 6.4\%  &  1.4\%  & 11.9\% & 2.6\%  &  19.0\% & 5.1\%    \\
    Jet veto                     & 4.0\%  &  0.8\%  & 7.4\% & 1.6\%  &  12.1\% & 3.1\% \\
    $p_{\rm T}(\ell\ell')> 30$ GeV     & 3.9\%  &  0.7\% & 7.1\% & 1.5\%  &  10.1\% & 2.6\% \\ \hline \hline
   \end{tabular}
   \caption{The product of acceptance times efficiency for the \WW\ simulated sample at each event selection step. 
            The $\tau\nu\ell\nu$ sample for the $ee$ channel includes both $WW \rightarrow \tau\nu e\nu$ 
            and $WW \rightarrow \tau\nu\tau\nu$ processes that result in two electrons in the final state; 
            and accordingly for the $\tau\nu\ell\nu$ samples for the $\mu\mu$ and $e\mu$ channels.}
   \label{ta:cutflow_mc_acc}
 \end{table*}

\begin{table*}[!ht]
\begin{center}
\begin{tabular}{cccc}
\hline \hline
 & \multicolumn{3}{c}{Relative uncertainty} \\
Source of uncertainty
           & $~~~~~~ee~~~~~~$ 
           & $~~~~~~\mu\mu~~~~~~$  
           & $~~~~~~e\mu~~~~~~$ 
\\
\hline 
PDFs                             & 0.9\%  & 0.9\% & 0.9\% \\
$\mu_R$ and $\mu_F$ scales       & 0.5\%  & 0.5\% & 0.6\% \\
Jet veto                         & 5.6\%  & 5.6\% & 5.6\% \\ 
\hline
Total                            & 5.7\%  & 5.7\% & 5.7\% \\
\hline \hline
\end{tabular}
\caption{\label{tbl:aww2}Relative uncertainties on the estimate of $A_{WW}$ 
for the $ee$, $\mu\mu$ and $e\mu$ channels.}
\end{center}
\end{table*}

\begin{table*}[!ht]
\centering
\begin{tabular}{cccc}
\hline \hline
 & \multicolumn{3}{c}{Relative uncertainty} \\
Source of uncertainty
           & $~~~~~~ee~~~~~~$ 
           & $~~~~~~\mu\mu~~~~~~$  
           & $~~~~~~e\mu~~~~~~$ 
\\
\hline 
Trigger efficiency               & 0.1\% & 0.6\% & 0.3\% \\
Lepton efficiency                & 2.9\% & 0.7\% & 1.4\% \\ 
Lepton \pt\ scale and resolution & 0.9\% & 0.8\% & 0.6\% \\ 
Jet energy scale and resolution  & 0.6\% & 0.5\% & 0.5\% \\ 
\met\ modeling                   & 0.5\% & 0.2\% & 0.4\% \\ 
Jet veto scale factor            & 2.8\% & 2.8\% & 2.7\% \\ 
PDFs, $\mu_R$ and $\mu_F$ scales & 0.7\% & 0.7\% & 0.3\% \\ 
\hline
Total                            & 4.2\% & 3.1\% & 3.2\% \\
\hline \hline
\end{tabular}
\caption{Relative uncertainties on the estimate of $C_{WW}$ 
for the $ee$, $\mu\mu$ and $e\mu$ channels.}
\label{tbl:cww}
\end{table*}

\begin{table*}[!ht]
\begin{center}
\begin{tabular}{cccc}
\hline \hline 
           & $~~~~~~ee~~~~~~$ 
           & $~~~~~~\mu\mu~~~~~~$  
           & $~~~~~~e\mu~~~~~~$ 
\\
\hline 
     $A_{WW}$    & $(7.5 \pm 0.1 \pm 0.4)\%$    & $(8.1 \pm 0.1 \pm 0.5)\%$  & $(15.9 \pm 0.1 \pm 0.9)\%$   \\
     $C_{WW}$    & $(40.3 \pm 0.5 \pm 1.7)\%$    & $(68.7 \pm 0.5 \pm 2.1)\%$  & $(50.5 \pm 0.2 \pm 1.6)\%$  \\
     $A_{WW} \times C_{WW}$    & $(3.0 \pm 0.1 \pm 0.1)\%$    & $(5.6 \pm 0.1 \pm 0.2)\%$  & $(8.0 \pm 0.1 \pm 0.3)\%$  \\
\hline \hline
\end{tabular}
\caption{\label{tbl:awwcww}
Acceptances $A_{WW}$, $C_{WW}$ and $A_{WW} \times C_{WW}$ for the $ee$, $\mu\mu$ and $e\mu$ channels.
           The first and second uncertainties represent the statistical and systematic uncertainties.} 
\end{center}
\end{table*}

\section{Background Estimation}
SM processes producing the $\ell\ell'+\met$ signature with no reconstructed jets in the final state are 
top-quark production, when additional jets in the final state are not reconstructed or identified (denoted by ``top-quark background"); 
$W$ production in association with jets (denoted by ``$W$+jets background") when one jet is reconstructed as a lepton; 
$Z/\gamma^\ast$ production in association with jets (denoted by ``Drell-Yan background") when apparent \met\ 
is generated from the mismeasurement of the \pt\ of the two 
leptons from $Z/\gamma^\ast$ boson decay; $WZ$ and $ZZ$ processes when only two leptons are reconstructed in the final state; 
and the $W\gamma$ process when the photon converts into electrons. 
The contribution from QCD multijet production when two jets are reconstructed as leptons is found to be negligible. 

\subsection{Background contribution from SM non-$WW$ diboson production processes}
The expected background contributions from SM non-$WW$ diboson processes ($WZ$, $ZZ$ and $W\gamma$) are estimated using simulation. 
The total number of selected non-$WW$ diboson background events corresponding to \nWWLumi~fb$^{-1}$ is estimated to be 
$13 \pm 1~({\rm stat}) \pm 2~({\rm syst})$, 
$21 \pm 1~({\rm stat}) \pm 2~({\rm syst})$, and 
$44 \pm 2~({\rm stat}) \pm 6~({\rm syst})$ 
for the $ee$, $\mu\mu$ and 
$e\mu$ channels, respectively. The systematic uncertainties arise mainly from theoretical uncertainties on the non-$WW$ diboson production 
cross sections and uncertainties on the lepton, jet and \met\ modeling in the simulation. 

\subsection{Background contribution from SM top-quark production processes}
Background contributions from top-quark production processes are suppressed by the jet veto requirement.
However, top-quark events containing no reconstructed jets with $p_{\rm T}>25$~GeV and $|\eta|<4.5$ could still mimic the signature of $WW$ candidates.
The top-quark background contribution is estimated using a data-driven method.

An extended signal region (ESR) is defined after the \metrel\ cut but before applying the jet veto and 
$p_{\rm T}(\ell\ell')$ criteria. In addition, a control region (CR) is defined as a subset of the ESR, which contains events 
having at least one $b$-tagged jet with $p_{\rm T}>20$~GeV. The jet multiplicity distribution for top-quark events 
in the ESR, $T^{\rm ESR}_{\rm data}$, is estimated from the jet multiplicity distribution in the CR, $T^{\rm CR}_{\rm data}$. 
In a first step, the non-top-quark background distribution $T^{\rm CR}_{\rm MC, nt}$ in the CR is estimated with simulation,   
scaled by a normalization factor $f_n^{\prime}$ and then subtracted from the measured $T^{\rm CR}_{\rm data}$ distribution. Subsequently, 
the resulting distribution is extrapolated bin-by-bin from the CR to the ESR via the MC prediction of the ratio 
$T^{\rm ESR}_{{\rm MC},i}/T^{\rm CR}_{{\rm MC},i}$ for each jet multiplicity bin $i$. The method can be summarized by the following equation for each jet multiplicity bin:
\begin{equation}
  T^{\rm ESR}_{\rm data} = \frac{T^{\rm ESR}_{\rm MC}}{T^{\rm CR}_{\rm MC}} (T^{\rm CR}_{\rm data} - f_n^{\prime} \times T^{\rm CR}_{\rm MC,nt}),
\end{equation}

\noindent where each symbol $T$ represents a full jet multiplicity distribution. The normalization scale 
factor $f_n^{\prime}$ for the non-top-quark background contributions in the CR is determined from events in the 
ESR by fitting the jet multiplicity distribution observed in data with the templates constructed from 
the data in the CR for top-quark contributions and from simulation for non-top-quark contributions. 
The value of $f_n^{\prime}$ is found to be $1.07 \pm 0.03$. In a final 
step, the number of top-quark background events in the signal region is estimated using the number of top-quark 
events in the ESR observed in data scaled by the ratio of top-quark events in the signal region to the number in the ESR 
in the MC simulation for the zero-jet bin.

The number of top-quark background events in the signal region is estimated to be 
$22 \pm 12~({\rm stat}) \pm 3~({\rm syst})$, 
$32 \pm 14~({\rm stat}) \pm 5~({\rm syst})$, and 
$87 \pm 23~({\rm stat}) \pm 13~({\rm syst})$ 
for the $ee$, $\mu\mu$ and $e\mu$ channels, respectively.
The statistical uncertainty is mainly due to the limited number of data events observed in the CR.
The systematic uncertainties are dominated by the $b$-tagging uncertainty.

An alternative data-driven method is used to cross-check the top-quark background estimation. 
To reduce the associated uncertainties on the jet veto probability, 
a data-based correction is derived from a top-quark dominated sample based on the 
\WW\ selection but with the requirement of at least one $b$-jet with $p_{\rm T}>25$~GeV~\cite{atlasww2}. 
In this sample, the ratio $P_1$ of events with one jet to the total number of events is sensitive to the modeling 
of the jet energy spectrum in top-quark events. A multiplicative correction based on the ratio 
$P_1^{\rm data}/P_1^{\rm MC}$ is applied to reduce the uncertainties resulting from the jet veto requirement. 
The results from the two data-driven methods are found to be consistent with each other within their uncertainties.

\subsection{Background contribution from $W$+jets production process}
The $W+$jets process can produce the $\ell\ell'+$\met\ signature when one jet is reconstructed as a charged lepton. 
Since the probability for a jet to be identified as a lepton may not be accurately modeled in the MC simulation,  
a data-driven method is employed to estimate this contribution. 
A leptonlike jet is defined as a jet that passes all lepton selection criteria but fails the 
lepton isolation requirement in the muon case, 
and fails at least one of the isolation or tight quality requirements in the electron case.
The ratio $f_\ell$ is then calculated as the ratio of jets satisfying the full lepton identification 
criteria to the number of leptonlike jets.
A jet-enriched data sample is selected containing one lepton that passes all lepton selection criteria and a leptonlike jet. 
The number of events in this sample is then scaled by the ratio $f_\ell$ to obtain the expected number of $W+$jets events in the signal region. 
The ratio $f_\ell$ is measured as a function of the jet \pt\ and \eta\ from a jet-enriched sample for electrons and muons separately. 
The number of $W$+jets background events in the signal regions is estimated to be 
$21 \pm 1~({\rm stat}) \pm 11~({\rm syst})$, 
$7 \pm 1~({\rm stat}) \pm 3~({\rm syst})$, and 
$70 \pm 2~({\rm stat}) \pm 31~({\rm syst})$ 
for the $ee$, $\mu\mu$ and 
$e\mu$ channels, respectively. 
The dominant source of systematic uncertainties stems from the $f_{\ell}$ measurement. 
The same method is applied to a $W$+jets-enriched sample selected with the requirement of two same-sign leptons to validate the $W$+jets estimation method. 
Consistent results are obtained for the number of observed and predicted events in this control region. 

An alternative method is used to check the $W$+jets estimation in the signal region.  
This method defines leptons with two different sets of quality criteria, one with the standard lepton selection criteria (called tight lepton here) 
and the other one with less restrictive lepton identification criteria (called loose lepton here). 
For loose muons, the isolation requirement is dropped. For loose electrons, the medium electron identification criteria as defined 
in Ref.~\cite{eid} are used and the isolation requirement is also dropped. 
Events with two loose leptons are assigned to one of four categories depending on 
whether both leptons, only the leading lepton, only the trailing lepton 
or neither of the two leptons, satisfy the tight lepton identification criteria. 
The corresponding numbers of events are denoted by $N_{TT}$, $N_{TL}$,  $N_{LT}$ and $N_{LL}$. 
The sample composition can be solved from a linear system of equations: 
\begin{equation}
\left (N_{TT}, N_{TL}, N_{LT}, N_{LL} \right)^T = {\cal{E}} \left( N_{\ell \ell'}, N_{\ell j}, N_{j \ell'}, N_{jj} \right)^T
\end{equation}

\noindent where $N_{\ell\ell'}$ is the number of events with two prompt leptons, $N_{\ell j}$ ($N_{j\ell'}$) is the number of 
events where only the leading (trailing) lepton is a prompt lepton, $N_{jj}$ is the number of events 
where neither of the two leptons are prompt leptons. 
The $4 \times 4$ matrix $\cal{E}$ contains the probabilities for a loose quality lepton to pass the tight quality selection for both prompt leptons and jets. 
These probabilities are estimated by applying the loose and tight selections to $Z/\gamma^* \rightarrow \ell\ell$ events and to a sample of 
dijet events, respectively. To take into account the lepton $p_{\rm T}$ dependence of these two probabilities, the matrix equation is inverted for 
each event, giving four weights, corresponding to these four combinations. These weights are then summed over 
all events in the signal region with loose lepton requirements to yield the estimated total number of 
background events from $W$+jets and dijet processes.
The results from the two data-driven methods are found to be consistent with each other within their uncertainties.

\subsection{Background contribution from Drell-Yan production process}
The Drell-Yan background is one of the dominant background contributions in the $ee$ and $\mu\mu$ channels. 
Its contribution is suppressed by the requirements on $m_{\ell\ell'}$, \metrel\ and $p_{\rm T}(\ell\ell')$. 
A control region dominated by the Drell-Yan process is defined by 
applying the same set of selection cuts as used for the signal region and reversing the $p_{\rm T}(\ell\ell')$ cut. 
The Drell-Yan background in the signal region is estimated
from the number of events observed in this control region,
after subtracting other background contributions using MC expectations,
scaled by the ratio of the number of MC $Z+$jets events in the signal region to the number in the control region.
The number of Drell-Yan background events in the signal region is estimated to be 
$12 \pm 3~({\rm stat}) \pm 3~({\rm syst})$, 
$34 \pm 6~({\rm stat}) \pm 10~({\rm syst})$ and 
$5 \pm 2~({\rm stat}) \pm 1~({\rm syst})$ 
events in the $ee$, $\mu\mu$ and $e\mu$ channels, respectively. 
As a cross-check, the results obtained above are compared to the predictions from simulation. Good agreement between the two estimates is found.

\section{Inclusive and Fiducial Cross-Section Results} 
Table~\ref{ta:selected_data_MC} shows the number of events selected in data and the estimated background 
contributions with statistical and systematic uncertainties for the three individual channels and the combined channel. 
The expected numbers of $WW$ signal events for the individual and the combined channels are also shown. 
In total \nWWNCandEvents~$\ell\ell'+\met$ candidates are observed in data with 
$824 \pm 4~({\rm stat}) \pm~69~({\rm syst})$ 
signal events expected from the $WW$ process and 
$369 \pm~31~({\rm stat}) \pm 53~({\rm syst})$ 
background events expected from non-$WW$ processes. 
The $WW$ processes mediated by a SM Higgs boson with a mass of 126~GeV would contribute an additional 3, 7 and 17 
events in the $ee$, $\mu\mu$ and $e\mu$ channels, respectively.
Figure~\ref{fig:candidates} shows the comparison between data and predictions for the leading lepton $p_{\rm T}$, 
azimuthal angle difference between the two leptons, $p_{\rm T}$ and the transverse mass $m_{\rm T}$ of the $\ell \ell'+\met$ system,   
where $m_{\rm T}$ is calculated as 
$\sqrt{ (E_{\rm T}^{\ell} + E_{\rm T}^{\ell'} + \met)^2 - (\vec{p}_{\rm T}^{~\ell} + \vec{p}_{\rm T}^{~\ell'} +\metvec)^2 }$ 
with $\vec{p}_{\rm T}^{~\ell}$ and $\vec{p}_{\rm T}^{~\ell'}$ being the transverse momentum vectors of the two leptons. 
The shapes of the Drell-Yan and top-quark distributions are taken from simulation and are scaled according to 
the data-driven estimates of the respective background. The $W$+jets background contribution is based on the data-driven method 
as described in Sec. VII C, and the non-$WW$ diboson background contributions are estimated using simulation.

The fiducial and total cross sections for the $WW$ process for the three individual decay channels are 
calculated using Eqs. (\ref{eqn:fidXsection}) and (\ref{eqn:totalXsection}), respectively. 
The results are shown in Table~\ref{ta:cross_section} together with the SM predictions. 
Reasonable agreement is found between the measured cross sections and the theoretical predictions.     
For the total cross-section measurement, the relative statistical uncertainty 
is 12\%, 8\% and 5\% for the $ee$, $\mu\mu$ and $e\mu$ channels, respectively, and the 
overall relative systematic uncertainty is 18\%, 10\% and 8\%, respectively. 

The combined total cross section from the three decay channels is determined by minimizing the 
negative log-likelihood function:
\begin{equation}
L = -\ln \prod^3_{i=1} \frac{e^{-(\mu^i_{\rm s}+\mu^i_{\rm b})} \times (\mu^i_{\rm s}+\mu^i_{\rm b})^{N^i_{\rm obs}}}{N^i_{\rm obs}!}
\label{eq:wwxsecfid}
\end{equation}

\noindent where $i=1, 2, 3$ runs over the three channels, $\mu^i_{\rm s}$ and $\mu^i_{\rm b}$ represent the 
expected $WW$ signal and estimated background for the \mbox{\rm $i$-th} channel, and $N^i_{\rm obs}$ represents the number of 
observed data events. The expected $WW$ 
signal is computed as $\mu^i_{\rm s} = \sigma_{WW} \times {\rm BR} \times {\cal L} \times A^i_{WW} \times C^i_{WW}$, where 
$A^i_{WW}$ and $C^i_{WW}$ are the corresponding $A_{WW}$ and $C_{WW}$ in the \mbox{\rm $i$-th} channel.

The combined total cross section is $\sigma_{WW} = \nWWCrossSection \pm \nWWCrossSectionStat~({\rm stat}) \pm  \nWWCrossSectionSys~({\rm syst}) \pm  \nWWCrossSectionLumi~({\rm lumi})$~pb and is also shown in Table~\ref{ta:cross_section}. 
The statistical uncertainty is estimated by taking the 
difference between the cross section at the minimum of the negative log-likelihood 
function and the cross section where the negative log-likelihood is 0.5 units above the minimum. 
Systematic uncertainties include all sources except luminosity and are taken into account by convolving the Poisson probability distributions 
for signal and background with the corresponding Gaussian distributions. Correlations between the signal and background uncertainties 
due to common sources of systematic uncertainties are taken into account in the definition of the likelihood. 

\begin{figure*}
  \begin{center}
    \subfigure[]{\includegraphics[width=0.49\textwidth]{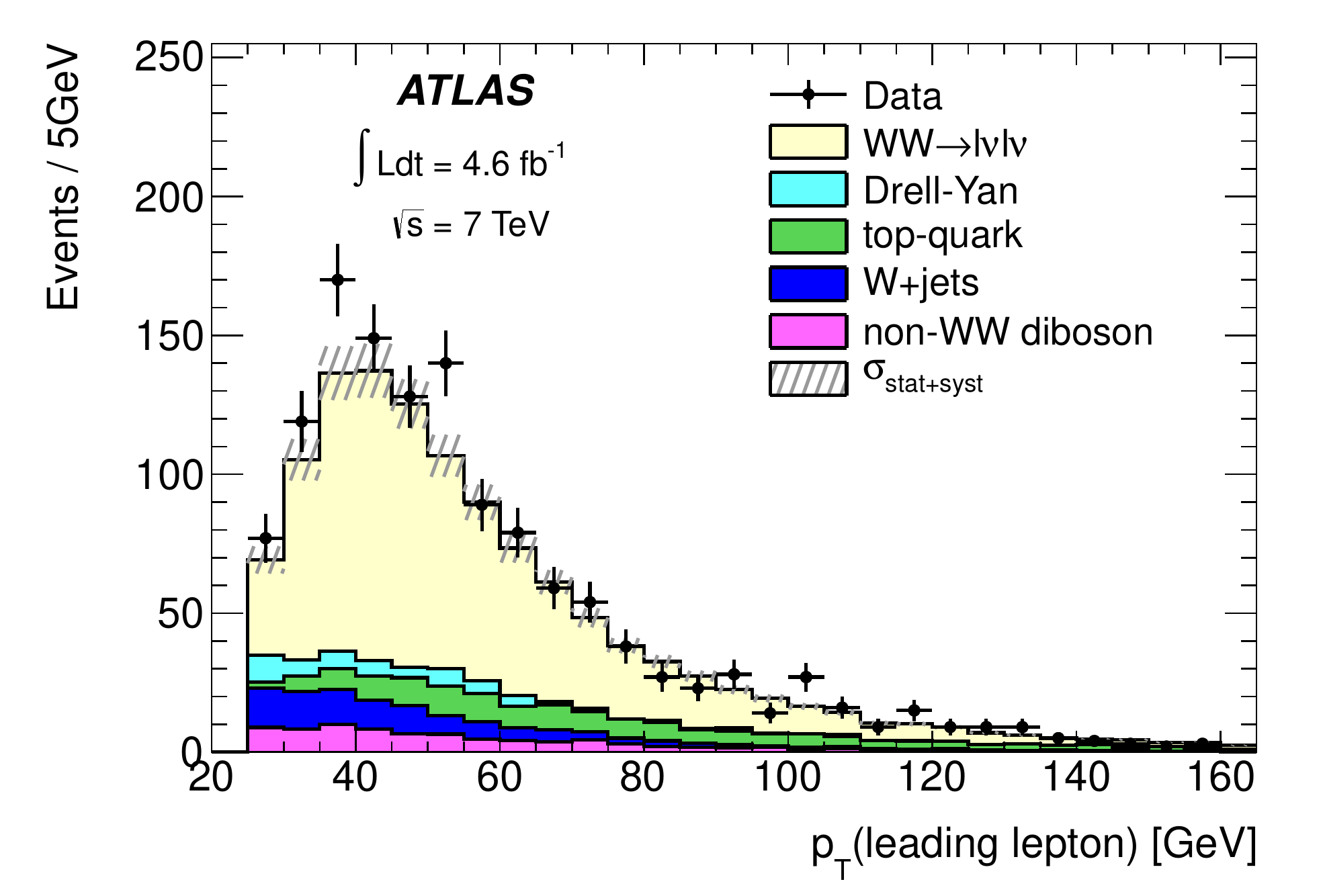}}
    \subfigure[]{\includegraphics[width=0.49\textwidth]{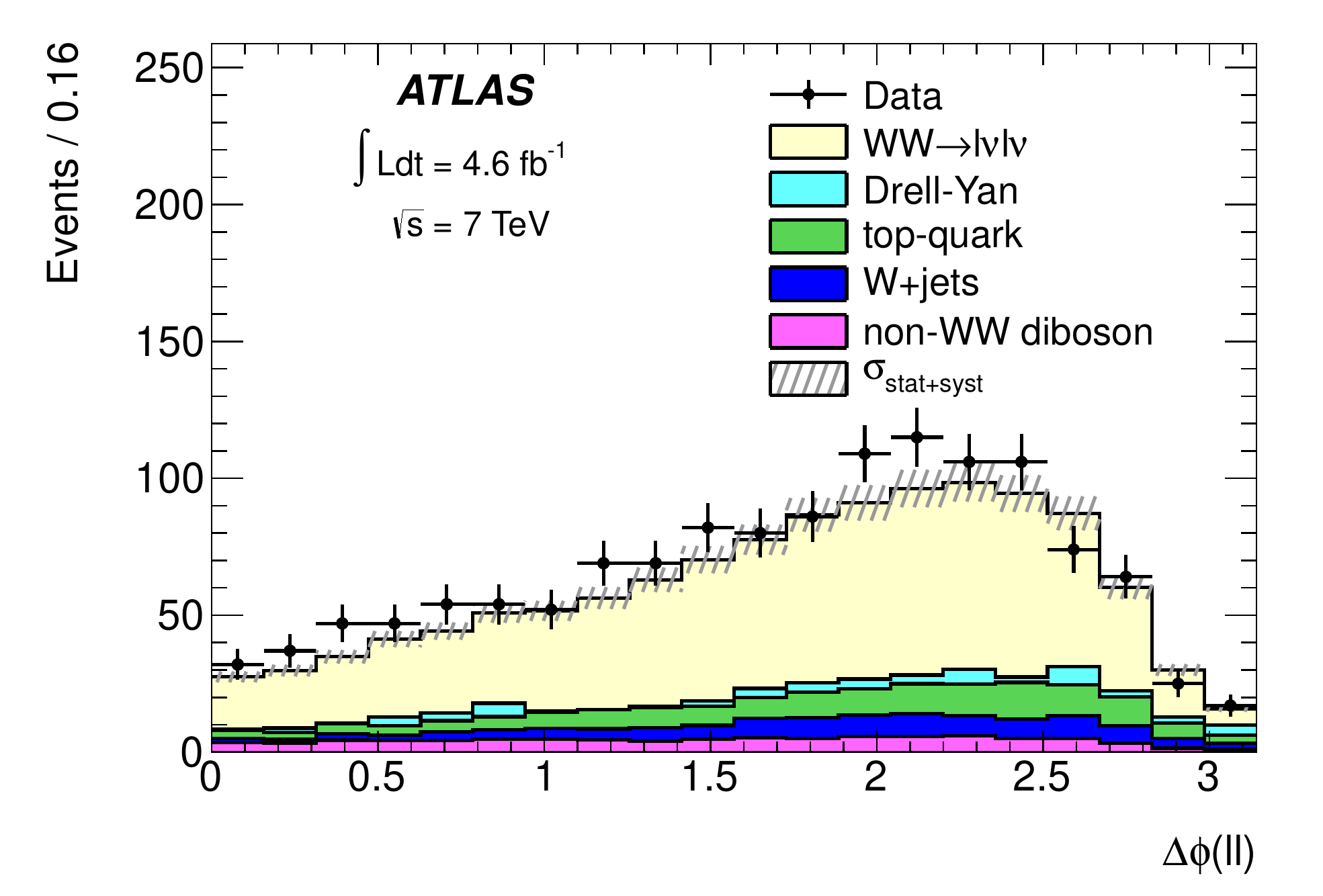}}\\ 
    \subfigure[]{\includegraphics[width=0.49\textwidth]{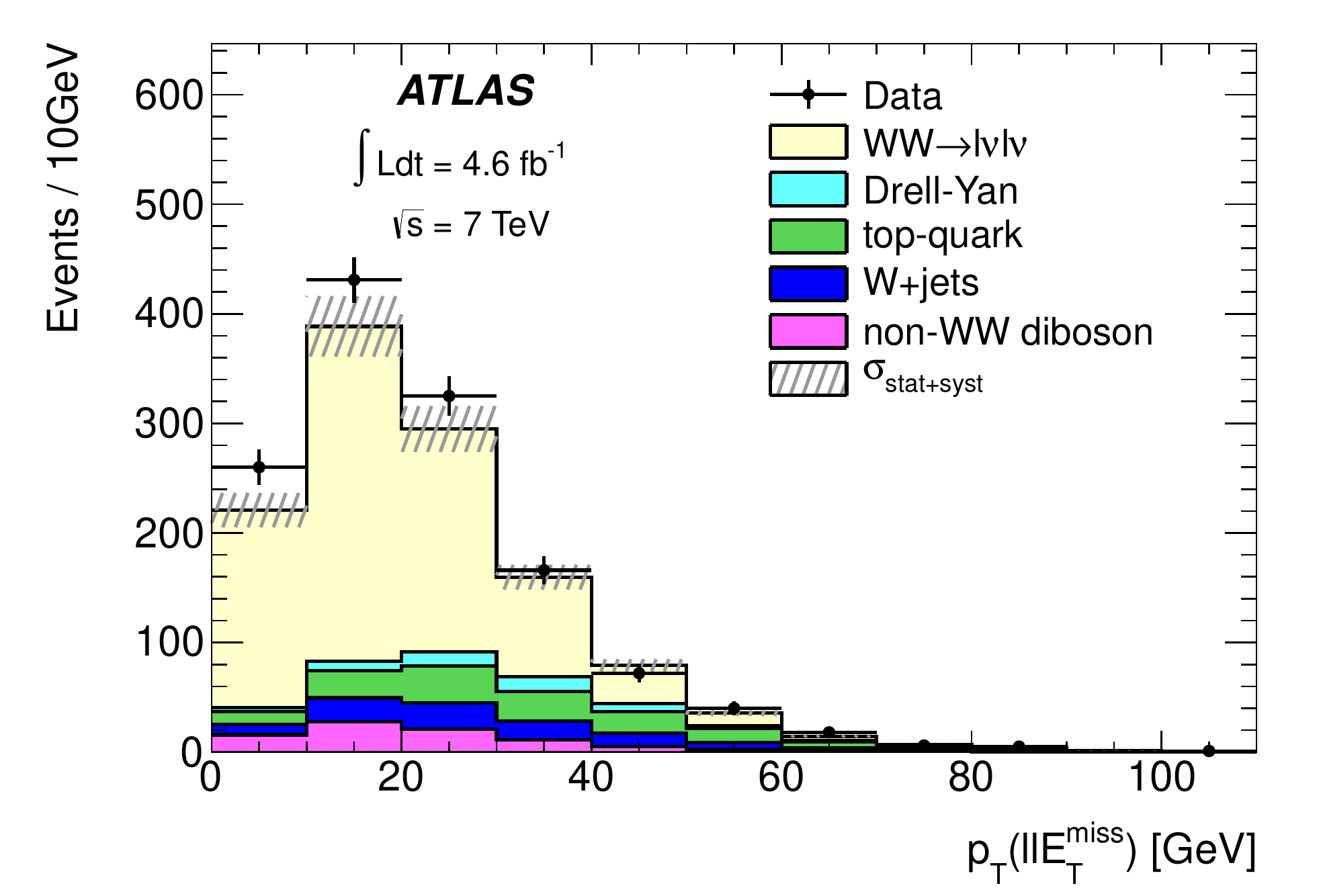}}
    \subfigure[]{\includegraphics[width=0.49\textwidth]{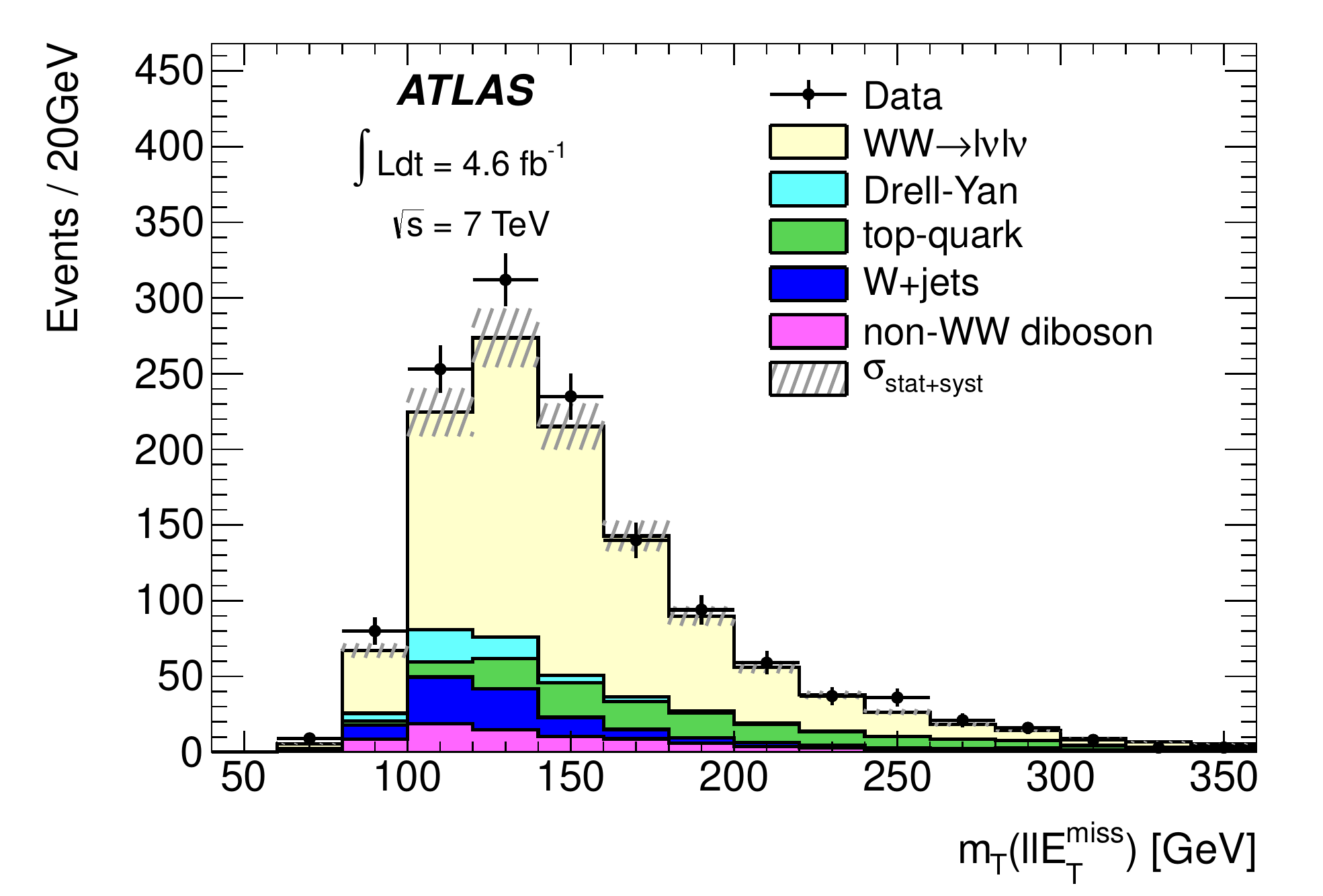}}
    \caption{Distributions for \WW\ candidates with all selection criteria applied and combining $ee$, $\mu\mu$ and $e\mu$ channels: 
      (a) leading lepton \pt\, (b) opening angle between the two leptons ($\Delta \phi(\ell \ell')$), 
      (c) $p_{\rm T}$ and (d) $m_{\rm T}$ of the $\ell\ell'+\met$ system. 
      The points represent data. 
      The statistical and systematic uncertainties are shown as grey bands. 
      The stacked histograms are from MC predictions except the background 
      contributions from the Drell-Yan, top-quark and $W$+jets processes, which are obtained from data-driven methods. 
      The prediction of the SM \WW\ contribution is normalized to the inclusive theoretical cross section of 44.7~pb.}
 \label{fig:candidates}
 \end{center}
 \end{figure*}

\begin{table*}
  \begin{tabular}{lcccc}
    \hline \hline
       & $~~~~~~~~~~ee~~~~~~~~~~$ & $~~~~~~~~~~~~~~~\mu\mu~~~~~~~~~~~~~~~$ & $~~~~~~~~~~~~~~~e\mu~~~~~~~~~~~~~~~$ & ~~~~~~~~~~Combined~~~~~~~~~~ \\ \hline
     Data  			& 174 				& 330 				& 821 					& 1325 \\ \hline
     $WW$             	& 100$\pm$2$\pm$9 	& 186$\pm$2$\pm$15  	& 538$\pm$3$\pm$45 		& 824$\pm$4$\pm$69   \\    \hline
     Top 		&22$\pm$12$\pm$3 	& 32$\pm$14$\pm$5 	& 87$\pm$23$\pm$13 		& 141$\pm$30$\pm$22 \\
     $W$+jets 		&21$\pm$1$\pm$11 	& 7$\pm$1$\pm$3 		& 70$\pm$2$\pm$31 		& 98$\pm$2$\pm$43 \\
     Drell-Yan 		&12$\pm$3$\pm$3 		& 34$\pm$6$\pm$10  	& 5$\pm$2$\pm$1 			& 51$\pm$7$\pm$12 \\
     Other dibosons 	&13$\pm$1$\pm$2 		& 21$\pm$1$\pm$2 		& 44$\pm$2$\pm$6 			& 78$\pm$2$\pm$10\\
     Total background	&68$\pm$12$\pm$13	& 94$\pm$15$\pm$13 	& 206$\pm$24$\pm$35 		& 369$\pm$31$\pm$53 \\ \hline
     Total expected   	&169$\pm$12$\pm$16 	& 280$\pm$16$\pm$20 	& 744$\pm$24$\pm$57 		& 1192$\pm$31$\pm$87 \\ 
    \hline \hline
  \end{tabular}
  \caption{\label{ta:selected_data_MC}\small Summary of observed and expected numbers of signal and background events
           in three individual channels and their combination (contributions from SM Higgs, VBF and DPS processes are not included). 
           The prediction of the SM \WW\ contribution is normalized to the inclusive theoretical cross section of 44.7~pb.   
           The first and second uncertainties represent the statistical and systematic uncertainties, respectively.} 
  
\end{table*}

\begin{table*}
  \begin{tabular}{lcccc}
    \hline \hline
             & Measured $\sigma^{\rm fid}_{WW}$ (fb)    & Predicted $\sigma^{\rm fid}_{WW}$ (fb)  & Measured $\sigma_{WW}$ (pb)  & Predicted $\sigma_{WW}$ (pb) \\ \hline
      $ee$   			& $56.4 \pm 6.8 \pm 9.8 \pm 2.2$       	& $54.6 \pm 3.7$            	& $46.9 \pm 5.7 \pm 8.2 \pm 1.8$       & $44.7^{+2.1}_{-1.9} $  \\           
      $\mu\mu$ 			& $73.9 \pm 5.9 \pm 6.9 \pm 2.9$  	& $58.9 \pm 4.0$            	& $56.7 \pm 4.5 \pm 5.5 \pm 2.2$       & $44.7^{+2.1}_{-1.9}$  \\           
      $e\mu$  			& $262.3 \pm 12.3 \pm 20.7 \pm 10.2$   	& $231.4 \pm 15.7$          	& $51.1 \pm 2.4 \pm 4.2 \pm 2.0$       & $44.7^{+2.1}_{-1.9}$  \\           
      Combined 			&   $\cdots$                		& $\cdots$             		& $51.9 \pm 2.0 \pm 3.9 \pm 2.0$       & $44.7^{+2.1}_{-1.9}$  \\           
    \hline \hline
  \end{tabular}
  \caption{\label{ta:cross_section}\small The measured fiducial and total cross sections for the 
                  three channels separately and also the total cross section for the combined channels, compared with theoretical predictions. 
                  The fiducial cross sections include the branching ratio for both $W$ bosons decaying into $e\nu$ 
                  or $\mu\nu$ (including decays through $\tau$ leptons with additional neutrinos).   
                  For the measured cross sections, the first uncertainty is statistical, the second is systematic 
                  without luminosity uncertainty and the third is the luminosity uncertainty.}
\end{table*}

\section{Normalized Differential Fiducial Cross Section} 
The measured leading lepton \pt\ distribution is unfolded to remove all experimental effects due 
to detector acceptance, resolution and lepton reconstruction efficiencies. The unfolded distribution 
provides a differential cross-section measurement in the fiducial phase space and allows a comparison 
with different theoretical models. A Bayesian unfolding technique~\cite{unfolding} with three iterative steps is used in this analysis. 

In unfolding of binned data, effects of the experimental acceptance and resolution are expressed in a response 
matrix, whose elements are the probability of an event in the $i$th bin at the generator level being reconstructed 
in the $j$th measured bin. 
The lepton \pt\ bins are chosen to be wider than the detector resolution to minimize migration effects and to maintain a sufficient number of 
events in each bin. The bin purity is found to be above 80\%, implying small bin-to-bin migration effects. 

The measured leading lepton \pt\ distribution in data is then corrected using a regularized inversion of the response matrix.
Finally, the distribution is corrected for efficiency and acceptance calculated from simulation.

Figure~\ref{fig:unfold_pt} shows the normalized fiducial cross sections 
($1/\sigfid \times d \sigfid/dp_{\rm T}$) extracted in bins of the leading 
lepton  \pt\ together with the SM predictions. The combined fiducial cross 
section $\sigfid$ is defined as the sum of the fiducial cross sections in each decay channel. 
The corresponding numerical values and the correlation matrix are shown in Table~\ref{tab:unfoldedInformation}. 
The overall uncertainty is about 5\% for leading lepton $\pt<80$~GeV and increases to 40\% for leading lepton $\pt>140$~GeV. 
The dominant source of uncertainty on the normalized differential cross section is statistical and 
is determined from MC ensembles. Two thousand pseudoexperimental spectra are generated by 
fluctuating the content of each bin according to a Poisson distribution with a mean that is equal to the bin content. The unfolding procedure 
is applied to each pseudoexperiment, and the root mean square of the results is taken as the statistical uncertainty. 

Systematic uncertainties on the normalized differential cross section mainly arise from uncertainties 
which directly impact the shape of the leading lepton \pt\ spectrum, i.e. the lepton energy scale and 
resolution, identification and isolation efficiencies, jet and \met\ modeling, and background 
estimations. The systematic uncertainties are evaluated by varying the response matrix for each 
uncertainty, and combining the resulting changes in the unfolded spectrum. 
Uncertainties on the expected background shapes and contributions are treated in a similar way. 
The performance of the unfolding procedure was verified by comparing the true and unfolded spectrum generated 
using pseudo-experiments. The unfolded results are stable with different numbers of iterations used and different 
input distributions. 

\begin{figure*}
  \begin{center}
    \includegraphics[width=0.6\textwidth]{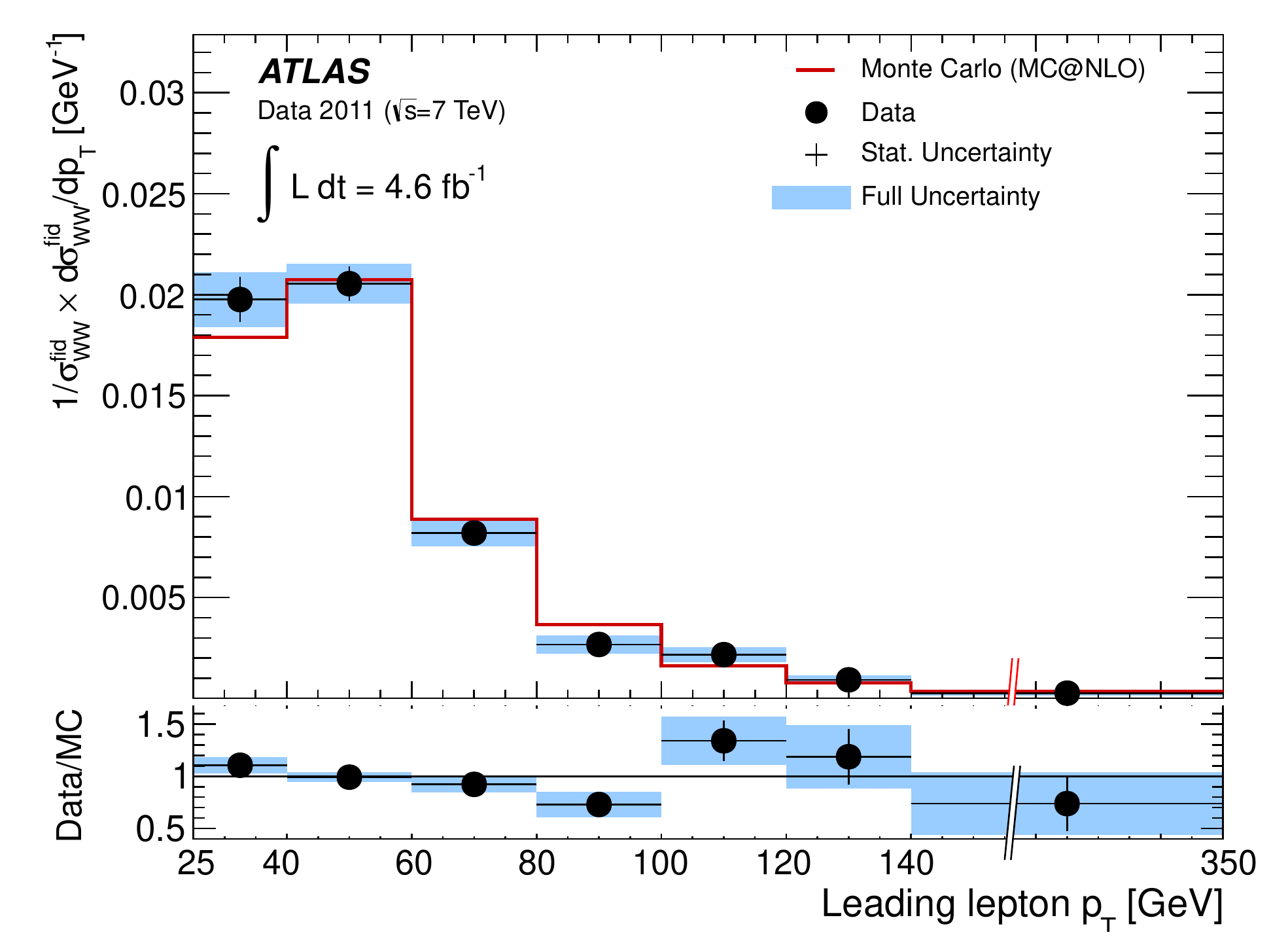}
    \caption{The normalized differential \WW\ fiducial cross section as a function of the leading lepton \pt\ compared to the SM prediction.}
 \label{fig:unfold_pt}
 \end{center}
 \end{figure*}
 
 \begin{table*}
  \centering
  \begin{tabular}{lccccccc} \hline \hline
Leading lepton \pt~[GeV]~~~~~	& ~~~[25,40]~~~     & ~~~[40,60]~~~		& 	~~~[60,80]~~~	& ~~~[80,100]~~~		& ~~~[100,120]~~~	& ~~~[120,140]~~~	& ~~~[140, 350]~~~	\\
Weighted bin center [GeV] 		& 33.6             & 50.2             & 70.2             & 89.1             & 107.1             & 127.5             & 180.4              \\
\hline
$1/\sigfid \times d \sigfid/dp_{\rm T}$ [GeV$^{-1}$] & $2.0 \times 10^{-2}$     & $2.1 \times 10^{-2}$    & $8.2 \times 10^{-3}$       & $2.7 \times 10^{-3}$	        & $2.2 \times 10^{-3}$		& $9.5 \times 10^{-4}$	& $6.2 \times 10^{-5}$ 	\\
Relative uncertainty				& 6.7\%	   & 4.8\%	     & 	8.2\%	& 17.0\%		& 17.1\%		& 25.5\%		& 41.0\% \\
\hline
Correlation	& 1      & $-$0.43  & $-$0.33  & $-$0.27  & $-$0.27  & $-$0.13  & $-$0.29  \\
&   & 1      & $-$0.29  & $-$0.29  & $-$0.23  & $-$0.30  & $-$0.15  \\
&   &   & 1      & $-$0.01  & $-$0.04  & 0.02  & 0.03  \\
&  &  &  & 1      & 0.21  & 0.11  & 0.14  \\
&  &  & &  & 1      & 0.23  & 0.11  \\
&  &  & &  &  & 1      & 0.27  \\
&  &  &  & &  &  & 1      \\
\hline
\hline
   \end{tabular}
   \caption{\label{tab:unfoldedInformation}Normalized fiducial cross section together with the overall uncertainty in bins of the leading lepton \pt. 
The weighted bin center is calculated as the cross-section-weighted average of the leading lepton \pt\ in each bin derived from {\sc mc@nlo} and {\tt gg2WW}.  
The correlation coefficients between 
different leading lepton \pt\ bins are also shown. Only half of the symmetric correlation matrix is presented.}
 \end{table*}
 
\section{Anomalous $WWZ$ and $WW\gamma$ couplings}
The reconstructed leading lepton $p_{\rm T}$ distribution is used to set limits on anomalous $WWZ$ and $WW\gamma$ TGCs.  
The Lorentz invariant Lagrangian describing the $WWZ$ and $WW\gamma$ interactions~\cite{lagrangian} has 14 independent coupling parameters. 
Assuming electromagnetic gauge invariance and C and P conservations,
the number of independent parameters reduces to five: $g_1^Z$, $\kappa_Z$, $\kappa_\gamma$, $\lambda_Z$ and $\lambda_\gamma$. 
In the SM, the coupling parameters have the following values: 
$g_1^Z=\kappa_Z=\kappa_\gamma=1$ and $\lambda_Z=\lambda_\gamma=0$. 
Deviations of these coupling parameters from their SM values  
$\Delta g_1^Z (\equiv g_1^Z - 1)$, 
$\Delta \kappa_Z (\equiv \kappa_Z - 1)$, $\Delta \kappa_\gamma (\equiv \kappa_\gamma-1)$, 
$\lambda_Z$ and $\lambda_\gamma$, all equal to zero in the SM, would result in an increase of the production
cross section and alter kinematic distributions, especially for large values of the leading lepton $p_{\rm T}$. 
Since unitarity restricts the $WWZ$ and $WW\gamma$ couplings to their SM values at asymptotically
high energies, each of the couplings is usually modified by $\alpha(\hat{s}) = \alpha_0/(1+\hat{s}/\Lambda^2)^2$, 
where $\alpha$ corresponds to one of the five couplings, $\alpha_0$ is the value of the anomalous 
coupling at low energy, $\hat{s}$ is the square of the invariant mass of the $WW$ system, 
and $\Lambda$ is the mass scale at which new physics affecting anomalous couplings would be introduced. 

Limits on these couplings can be obtained under the assumption that the $WWZ$ and $WW\gamma$ couplings are equal 
(denoted by the ``equal couplings scenario'') ($\Delta \kappa_Z = \Delta \kappa_\gamma$, $\lambda_Z = \lambda_\gamma$, and $g_1^Z=1$). 
Two other different sets of parameters are also considered. One, motivated by $SU(2) \times U(1)$ gauge invariance, was 
used by the LEP collaborations (denoted by the ``LEP scenario'')~\cite{lepatgc} and assumes 
$\Delta \kappa_\gamma = (\cos^2\theta_W/\sin^2\theta_W)(\Delta g_1^Z - \Delta\kappa_Z)$, and $\lambda_Z = \lambda_\gamma$. 
The other one (denoted by the ``HISZ scenario'')~\cite{hisz} assumes 
$\Delta g_1^Z = \Delta\kappa_Z / (\cos^2\theta_W-\sin^2\theta_W)$, 
$\Delta\kappa_\gamma = 2 \Delta\kappa_Z \ \cos^2\theta_W/(\cos^2\theta_W-\sin^2\theta_W)$, and $\lambda_Z = \lambda_\gamma$.
Due to the constraints mentioned above, the number of independent parameters is only two for the Equal Couplings scenario and the HISZ scenario, 
and three for the LEP scenario. Limits are also set assuming no relationships among these five parameters.  

A reweighting method is applied to SM $WW$ events generated with {\sc mc@nlo} and processed through 
the full detector simulation to obtain the leading lepton $p_{\rm T}$ distribution 
with anomalous couplings. The reweighting method uses an event weight to predict the rate with which a given 
event would be generated if anomalous couplings were present. 
The event weight is the ratio of the squared matrix elements with and without anomalous couplings i.e.,
$|{\cal M}|^2 /|{\cal M}|^2_{\rm SM}$, where $|{\cal M}|^2$ is the matrix element squared in the presence of 
anomalous couplings and $|{\cal M}|^2_{\rm SM}$ is the matrix element squared in the SM. 
The event generator {\sc bho}~\cite{bho} is used for the calculation of the two matrix elements. 
Generator-level comparisons of \WW\ production between {\sc mc@nlo} and {\sc bho} with all anomalous 
couplings set to zero are performed and consistent results are obtained.
Samples with different sets of anomalous couplings are generated and the ratio of the leading lepton $p_{\rm T}$ distribution to the SM prediction 
is parameterized as a function of the input anomalous coupling parameters. This function is then used to interpolate 
the leading lepton $p_{\rm T}$ distribution for any given anomalous couplings. 
To verify the reweighting method, the event weights for a given set of anomalous couplings are calculated and applied to events 
generated with {\sc bho} assuming no anomalous couplings. The reweighted distributions are compared to those predicted by 
the {\sc bho} generator, and good agreement is observed for the inclusive cross section and for the kinematic 
distributions as shown in Fig.~\ref{fig:reweight_compare}(a).
 
Figure~\ref{fig:reweight_compare}(b) compares the reconstructed leading lepton $p_{\rm T}$ spectrum in data with 
that from the sum of expected signal and background contributions. 
The predicted leading lepton $p_{\rm T}$ distributions for three different anomalous TGC values are also shown.  
Events at high values of the leading lepton $p_{\rm T}$ distribution are sensitive to anomalous TGCs.  
Limits on anomalous TGCs are obtained by forming a likelihood test incorporating the observed 
number of candidate events, the expected signal as a function of anomalous TGCs and the estimated 
number of background events in each $p_{\rm T}$ bin. The systematic uncertainties are 
included in the likelihood function as nuisance parameters with correlations taken into account. 
The 95\% confidence level (C.L.) intervals on anomalous TGC parameters include all values of anomalous TGC parameters 
for which the negative log-likelihood functions increase by no more than 1.92 (2.99) 
units above the minimum for the one (two)-dimensional case. 

Table~\ref{ta:aTGC_results} shows expected and observed 95\% C.L. limits on 
anomalous $WWZ$ and $WW\gamma$ couplings for three scenarios (LEP, HISZ and equal couplings)
with two scales, $\Lambda=6$~TeV and $\Lambda=\infty$. The $\Lambda=6$~TeV scale is chosen as it is the rounded largest 
value which still preserves unitarity for all extracted anomalous TGC limits of this analysis. 
Table~\ref{ta:aTGC_results2} shows the results assuming no relationships between the five couplings.
Figure~\ref{fig:aTGC_2D} shows the two-dimensional 95\% C.L. contour limits of $\Delta \kappa_Z$ vs. $\lambda_Z$, $\Delta \kappa_Z$ vs. $\Delta g^Z_1$, 
$\Delta \kappa_\gamma$ vs. $\Delta g_1^Z$ and $\lambda_Z$ vs. $\Delta g_1^Z$ for the LEP scenario.
Except for the anomalous coupling parameter(s) under study, all other parameters are set to their SM values.

Limits in the LEP scenario are compared with limits obtained from the CMS~\cite{cmsww}, CDF~\cite{tev}, 
D\O~\cite{tev} and LEP~\cite{lep} experiments in Fig.~\ref{fig:aTGC_compare}. 
Due to higher energy and higher \WW\ production cross section at the LHC, the limits obtained in this paper are better than 
the Tevatron results and approach the precision of the combined limits from the LEP experiments. 

\begin{table*}
\centering
\begin{tabular}{l c c ccc}
\hline\hline
           						 &     							& Expected  	&  Observed   	& Expected  		& Observed \\
\raisebox{1.ex} {Scenario}  & \raisebox{1.ex}{~~~~~Parameter~~~~~}  & ~~~($\Lambda=6$ TeV)~~~  &  ~~~($\Lambda=6$ TeV)~~~   & ~~~~~($\Lambda=\infty$)~~~~~ & ~~~~~($\Lambda=\infty$)~~~~~ \\ [0.5ex]
\hline
 						&  $\Delta \kappa_Z$    & [$-$0.043, 0.040] & [$-$0.045, 0.044] & [$-$0.039, 0.039] & [$-$0.043, 0.043] \\
LEP   						&  $\lambda_Z=\lambda_\gamma$   & [$-$0.060, 0.062] & [$-$0.062, 0.065] & [$-$0.060, 0.056] & [$-$0.062, 0.059] \\
 						&  $\Delta g_1^Z$      	& [$-$0.034, 0.062] & [$-$0.036, 0.066] & [$-$0.038, 0.047] & [$-$0.039, 0.052]\\ \hline
						&  $\Delta \kappa_Z$    & [$-$0.040, 0.054] & [$-$0.039, 0.057] & [$-$0.037, 0.054] & [$-$0.036, 0.057] \\ 
\raisebox{1.ex}{HISZ} 			&  $\lambda_Z=\lambda_\gamma$   & [$-$0.064, 0.062] & [$-$0.066, 0.065] & [$-$0.061, 0.060] & [$-$0.063, 0.063] \\ \hline
						&  $\Delta \kappa_Z$  	& [$-$0.058, 0.089] & [$-$0.061, 0.093] & [$-$0.057, 0.080] & [$-$0.061, 0.083] \\ 
\raisebox{1.ex}{Equal Couplings} 	&  $\lambda_Z=\lambda_\gamma$   & [$-$0.060, 0.062] & [$-$0.062, 0.065] & [$-$0.060, 0.056] & [$-$0.062, 0.059] \\ \hline
\hline
\end{tabular}
  \caption{\label{ta:aTGC_results}The 95\% C.L. expected and observed limits on anomalous TGCs in 
           the LEP, HISZ and Equal Couplings scenarios. Except for the coupling under study, 
           all other anomalous couplings are set to zero. 
           The results are shown for two scales $\Lambda=6$~TeV and $\Lambda=\infty$.}
\end{table*}

\begin{table*}
\centering
\begin{tabular}{ccc}
\hline\hline
             & Expected  & Observed \\
 \raisebox{1.ex}{~~~~~Parameter~~~~~}  & ~~~~~($\Lambda=\infty$)~~~~~ & ~~~~~($\Lambda=\infty$)~~~~~ \\ [0.5ex]
\hline
      $\Delta \kappa_Z$            & [$-$0.077, 0.086] & [$-$0.078, 0.092] \\ 
      $\lambda_Z$                  & [$-$0.071, 0.069] & [$-$0.074, 0.073] \\ 
      $\lambda_\gamma$      & [$-$0.144, 0.135] & [$-$0.152, 0.146] \\ 
      $\Delta g_1^Z$               & [$-$0.449, 0.546] & [$-$0.373, 0.562] \\ 
      $\Delta \kappa_\gamma$       & [$-$0.128, 0.176] & [$-$0.135, 0.190] \\ 
\hline
\hline
\end{tabular}
  \caption{\label{ta:aTGC_results2}The 95\% C.L. expected and observed limits on anomalous TGCs assuming no relationships between these five coupling parameters for $\Lambda=\infty$. Except for the coupling under study, all other anomalous couplings are set to zero.}
  
\end{table*}

\begin{figure*}
  \begin{center}
    \includegraphics[width=0.45\textwidth]{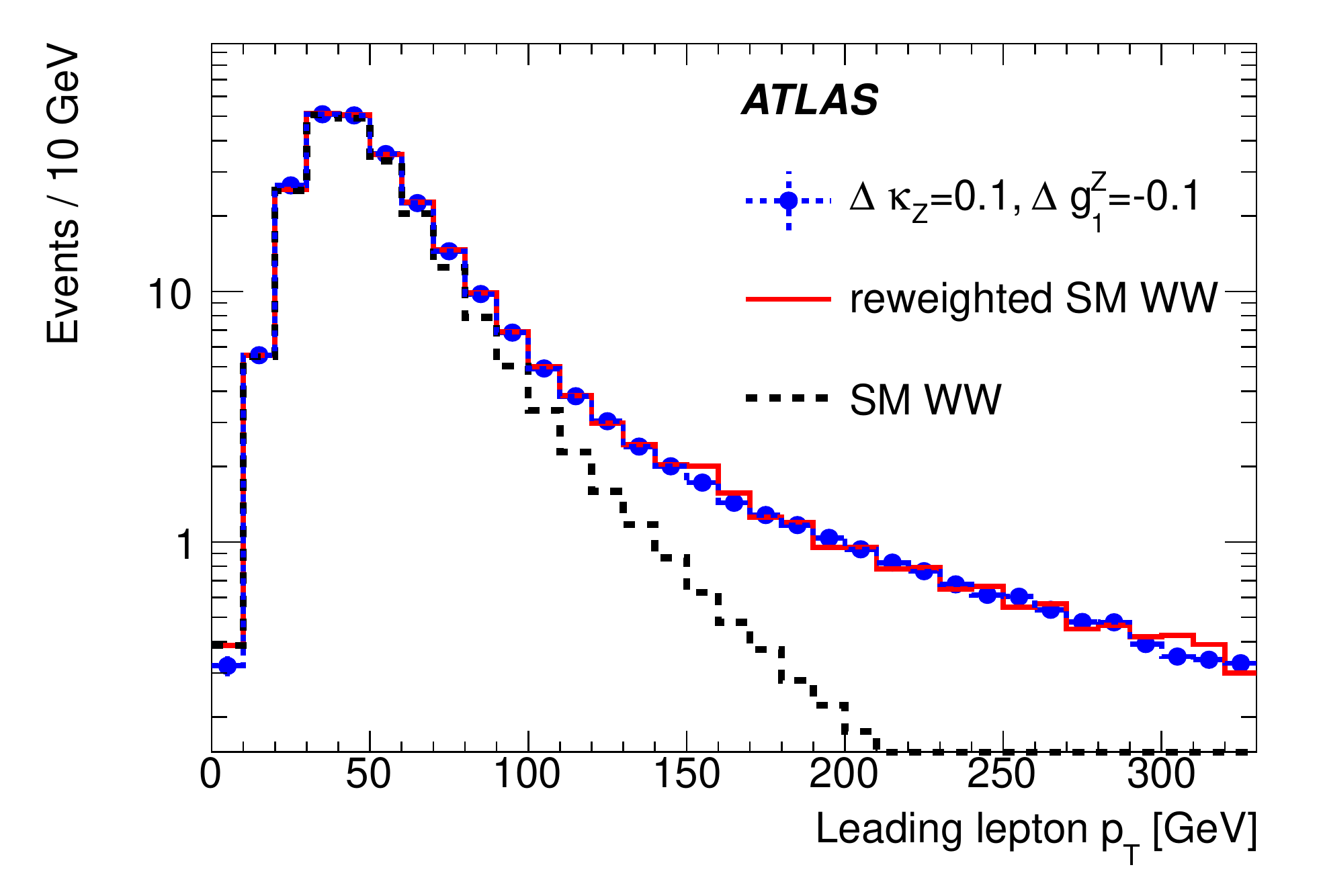}  \hspace{-4cm}{(a)}\hspace{4cm}
    \includegraphics[width=0.45\textwidth]{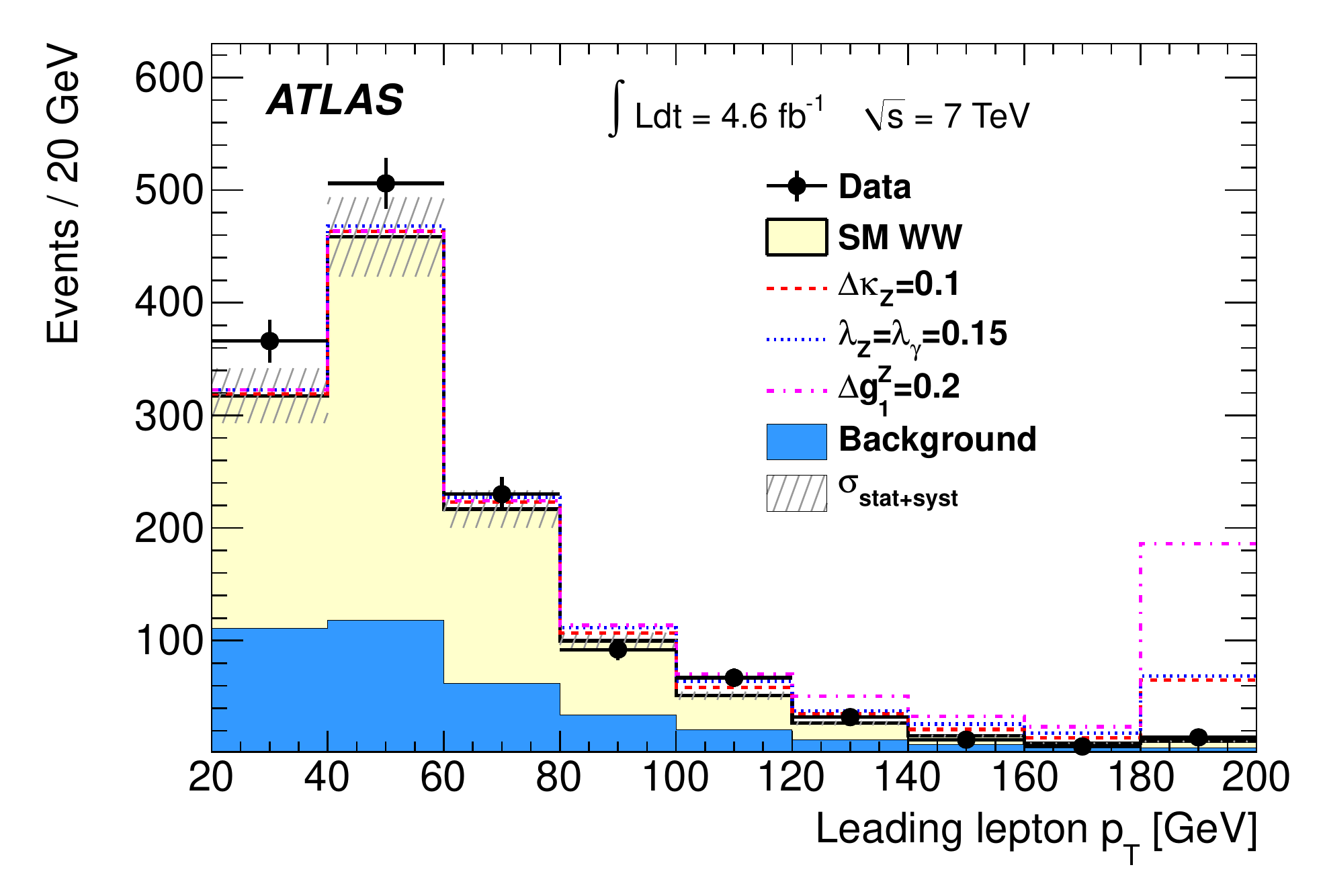} \hspace{-4cm}{(b)}\hspace{4cm}
    \caption{(a) The leading lepton $p_{\rm T}$ spectrum from the SM prediction, compared with a prediction using {\sc bho} 
             and by reweighting the SM prediction assuming the LEP scenario with $\Delta \kappa_Z=0.1$, $\lambda_Z=0$, $\Delta g_1^Z=-0.1$ 
             and $\Lambda=\infty$; (b) The reconstructed leading lepton $p_{\rm T}$ spectrum in data and sum of MC signal and background for the SM prediction 
             and for three different anomalous TGC predictions. The shaded band corresponds to the total statistical and systematic uncertainties. 
             The rightmost bin shows the sum of all events with leading lepton $p_{\rm T}$ above 180~GeV.}
 \label{fig:reweight_compare}
 \end{center}
 \end{figure*}

\begin{figure*}
  \begin{center}
   \subfigure[]{\includegraphics[width=0.4\textwidth]{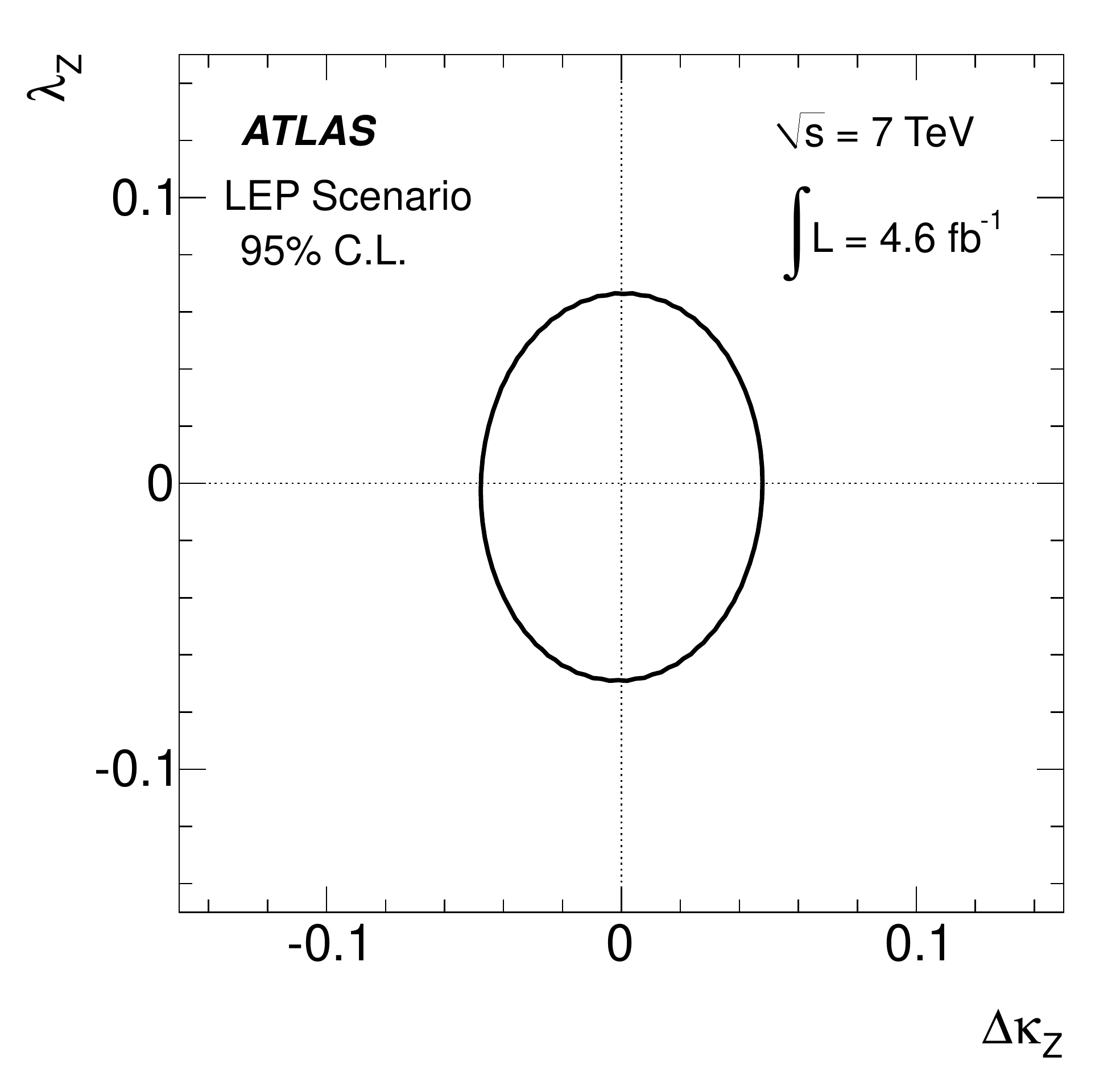}}
   \subfigure[]{\includegraphics[width=0.4\textwidth]{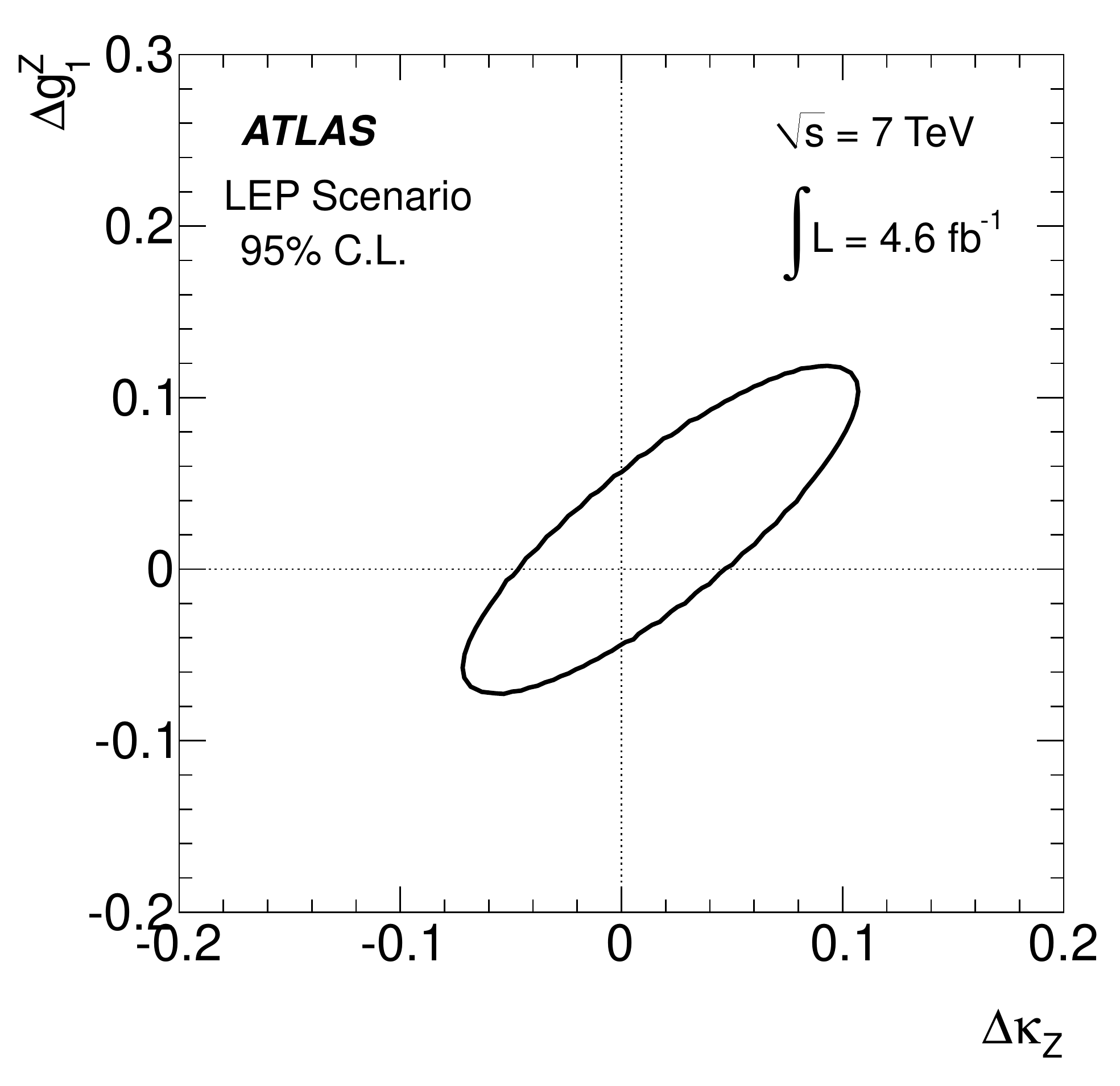}}\\
   \subfigure[]{\includegraphics[width=0.4\textwidth]{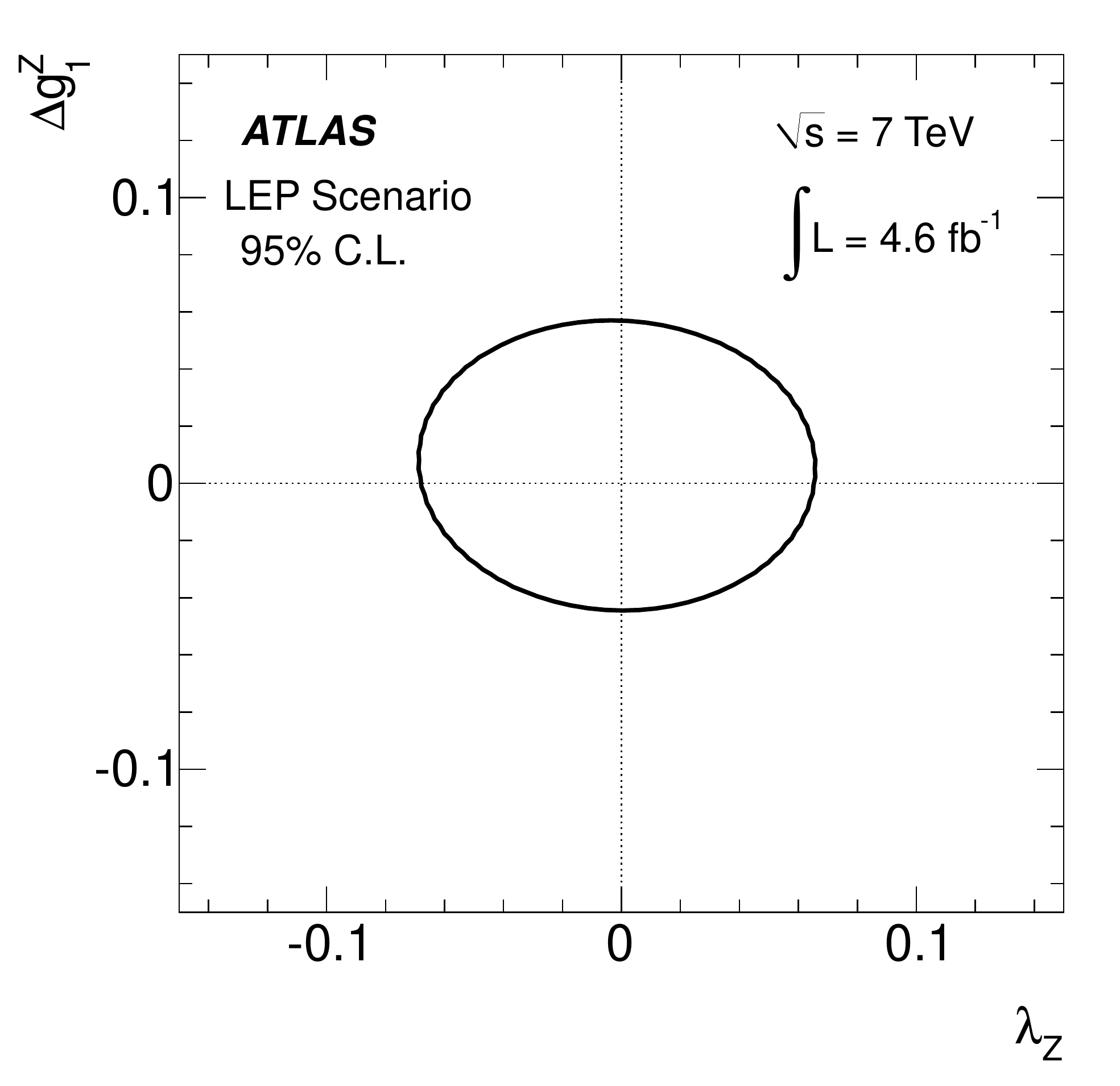}}
   \subfigure[]{\includegraphics[width=0.4\textwidth]{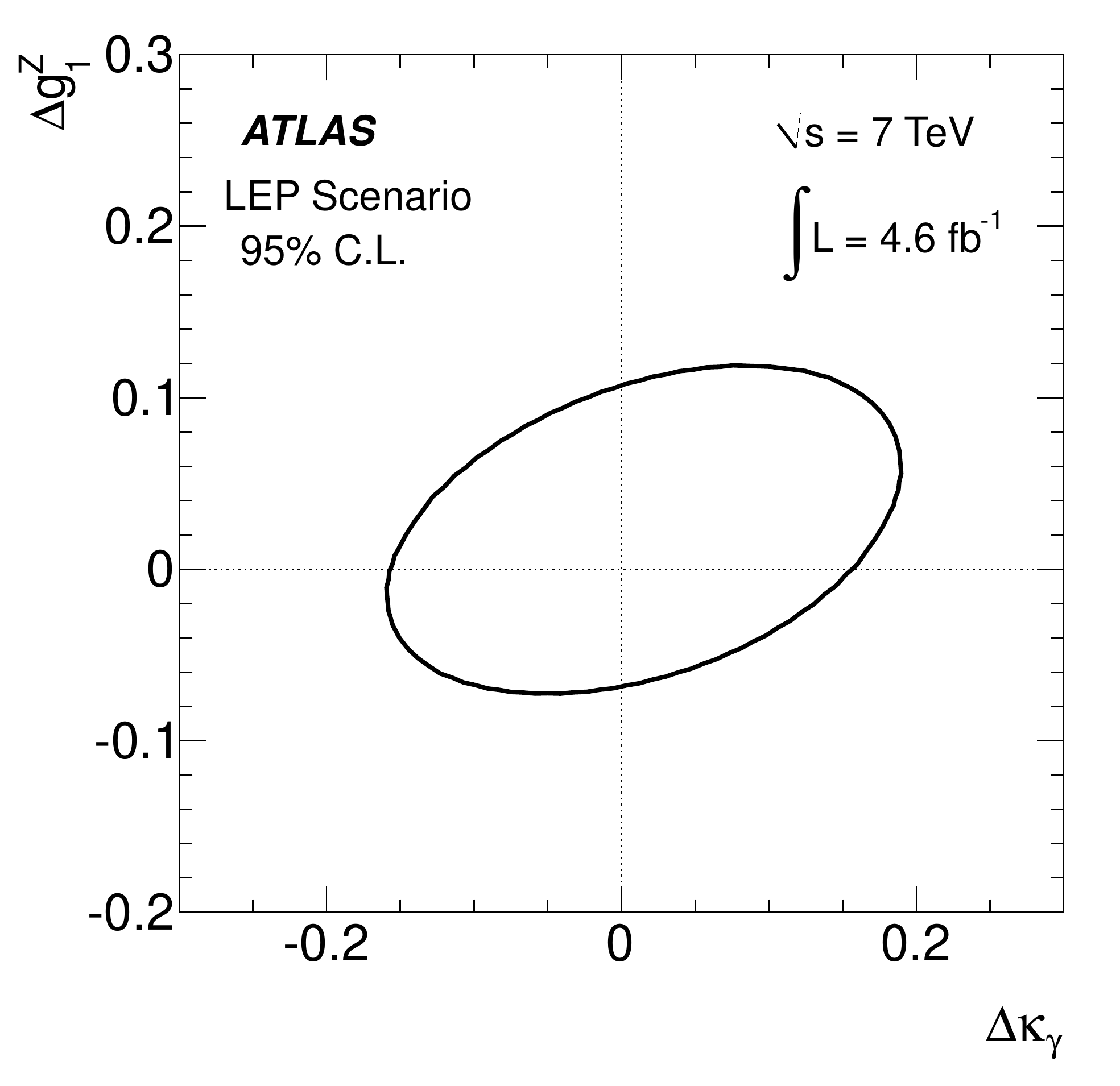}}
    \caption{Two-dimensional 95\% C.L. contour limits on (a) $\lambda_Z$ vs $\Delta \kappa_Z$, 
             (b) $\Delta g^Z_1$ vs $\Delta \kappa_Z$, 
             (c) $\Delta g_1^Z$ vs $\lambda_Z$ and 
             (d) $\Delta g^Z_1$ vs $\Delta \kappa_\gamma$ for the LEP scenario for $\Lambda=\infty$. 
             Except for the two parameters under study, all other anomalous couplings are set to zero.}
 \label{fig:aTGC_2D}
 \end{center}
 \end{figure*}

\begin{figure*}
  \begin{center}
    \includegraphics[width=0.6\textwidth]{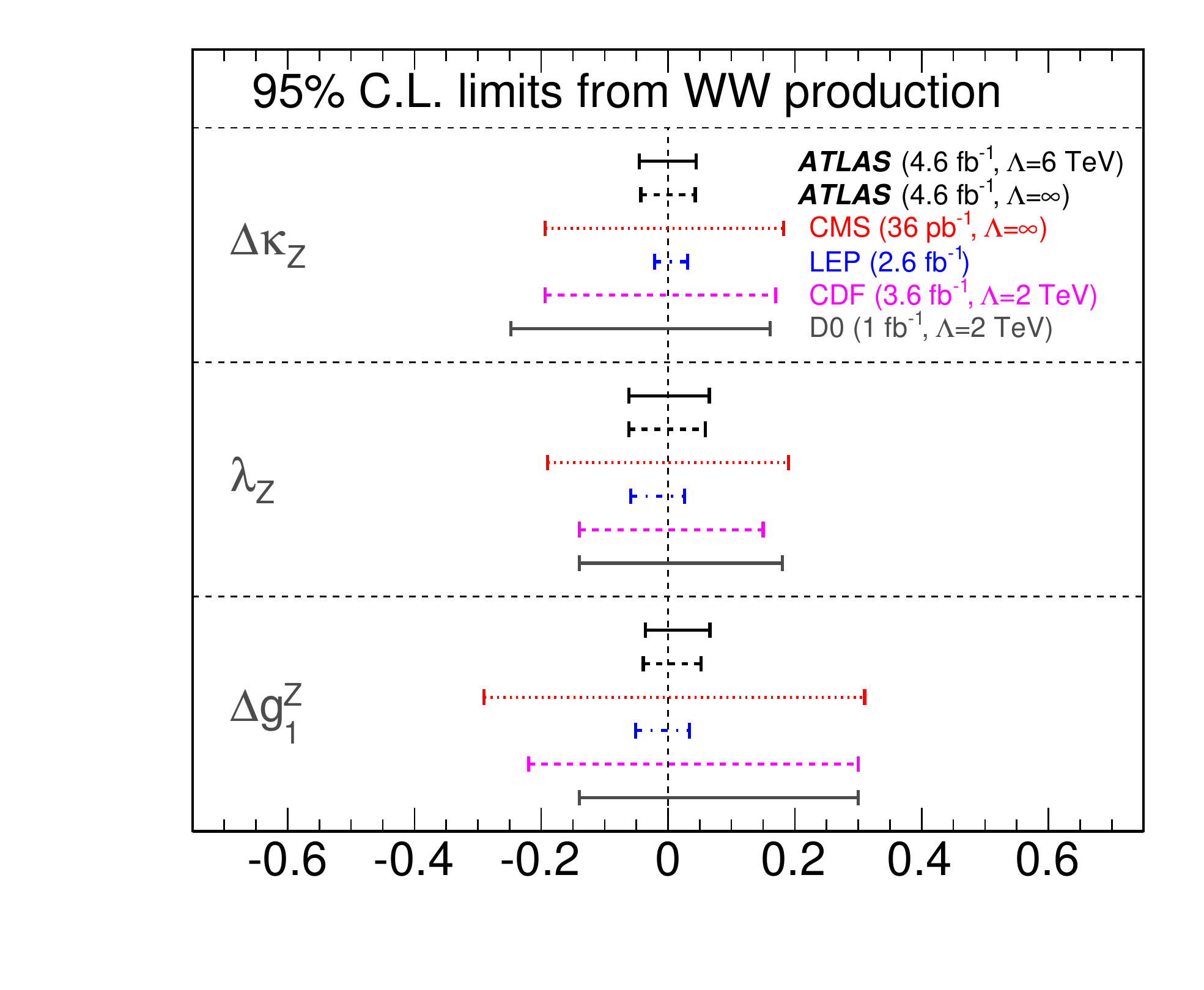}
    \caption{Comparison of anomalous TGC limits from ATLAS, CMS, CDF, D\O\ and LEP experiments for the LEP scenario.
    The $\Delta \kappa_Z$ result in the LEP scenario from CMS was obtained using the $\Delta \kappa_\gamma$ limit 
    in the HISZ scenario \cite{cmsww} and inserting it in the LEP scenario assuming $\Delta g^Z_1 = 0$.}
 \label{fig:aTGC_compare}
 \end{center}
 \end{figure*}

\section{Conclusion} 
The \WW\ production cross section in $pp$ collisions at $\sqrt s = 7$~TeV is measured using \nWWLumi\ \ifb\ 
of data collected with the ATLAS detector at the LHC.  The measurement is conducted using the $WW \rightarrow \ell\nu\ell' \nu'$ ($\ell, \ell'=e, \mu$) channels 
including decays through $\tau$ leptons with additional neutrinos.  
In total \nWWNCandEvents~candidates are selected with
an estimated background of $\nWWNBGEvents \pm \nWWNBGEventsUncert$ events for the three decay channels into $ee$, $\mu\mu$ and $e\mu$ final states. 
The combined production cross section $\sigma(pp\rightarrow WW+X)$ is 
$\nWWCrossSection \pm \nWWCrossSectionStat~{\rm (stat)} \pm
\nWWCrossSectionSys~{\rm (syst)} \pm \nWWCrossSectionLumi~{\rm (lumi)}$~pb, 
compatible with the SM NLO prediction of \nWWSMCrossSectionNLO \nWWSMCrossSectionUncertNLO~pb.
The overall statistical and systematic uncertainty is 9\% and an improvement of 30\% has been 
achieved compared with the previous ATLAS measurement~\cite{atlasww2}. 
The results presented supersede the previous results obtained with 1 fb$^{-1}$ of data.  
Cross sections are also measured in a fiducial phase space. 

The leading lepton $p_{\rm T}$ distribution is unfolded to obtain the normalized differential fiducial 
cross section in the chosen fiducial phase space. Reasonable agreement is observed between 
the measured distribution and theoretical predictions using {\sc mc@nlo}.

Anomalous $WWZ$ and $WW\gamma$ couplings are probed using the reconstructed leading lepton $p_{\rm T}$ distribution of the selected $WW$ events. 
With the assumption that $WWZ$ and $WW\gamma$ couplings are equal, 95\% C.L. limits are set on $\Delta\kappa_Z$ and $\lambda_Z$ in the intervals
$[-0.061, 0.093]$ and $[-0.062, 0.065]$ respectively for a scale of $\Lambda = 6$~TeV.
Limits on these anomalous couplings are also reported for three other scenarios and two scales $\Lambda=6$ TeV and $\Lambda=\infty$. 
The limits on anomalous TGCs obtained approach the precision of the combined limits from the four LEP experiments. 

\section{Acknowledgments}

We thank CERN for the very successful operation of the LHC, as well as the
support staff from our institutions without whom ATLAS could not be
operated efficiently.

We acknowledge the support of ANPCyT, Argentina; YerPhI, Armenia; ARC,
Australia; BMWF and FWF, Austria; ANAS, Azerbaijan; SSTC, Belarus; CNPq and FAPESP,
Brazil; NSERC, NRC and CFI, Canada; CERN; CONICYT, Chile; CAS, MOST and NSFC,
China; COLCIENCIAS, Colombia; MSMT CR, MPO CR and VSC CR, Czech Republic;
DNRF, DNSRC and Lundbeck Foundation, Denmark; EPLANET and ERC, European Union;
IN2P3-CNRS, CEA-DSM/IRFU, France; GNSF, Georgia; BMBF, DFG, HGF, MPG and AvH
Foundation, Germany; GSRT, Greece; ISF, MINERVA, GIF, DIP and Benoziyo Center,
Israel; INFN, Italy; MEXT and JSPS, Japan; CNRST, Morocco; FOM and NWO,
Netherlands; BRF and RCN, Norway; MNiSW, Poland; GRICES and FCT, Portugal; MERYS
(MECTS), Romania; MES of Russia and ROSATOM, Russian Federation; JINR; MSTD,
Serbia; MSSR, Slovakia; ARRS and MVZT, Slovenia; DST/NRF, South Africa;
MICINN, Spain; SRC and Wallenberg Foundation, Sweden; SER, SNSF and Cantons of
Bern and Geneva, Switzerland; NSC, Taiwan; TAEK, Turkey; STFC, the Royal
Society and Leverhulme Trust, United Kingdom; DOE and NSF, United States of
America.

The crucial computing support from all WLCG partners is acknowledged
gratefully, in particular from CERN and the ATLAS Tier-1 facilities at
TRIUMF (Canada), NDGF (Denmark, Norway, Sweden), CC-IN2P3 (France),
KIT/GridKA (Germany), INFN-CNAF (Italy), NL-T1 (Netherlands), PIC (Spain),
ASGC (Taiwan), RAL (UK) and BNL (USA) and in the Tier-2 facilities
worldwide.

\onecolumngrid
\clearpage
\begin{flushleft}
{\Large The ATLAS Collaboration}

\bigskip

G.~Aad$^{\rm 48}$,
T.~Abajyan$^{\rm 21}$,
B.~Abbott$^{\rm 111}$,
J.~Abdallah$^{\rm 12}$,
S.~Abdel~Khalek$^{\rm 115}$,
A.A.~Abdelalim$^{\rm 49}$,
O.~Abdinov$^{\rm 11}$,
R.~Aben$^{\rm 105}$,
B.~Abi$^{\rm 112}$,
M.~Abolins$^{\rm 88}$
% ...
\bigskip
\\
% ...
$^{11}$ Institute of Physics, Azerbaijan Academy of Sciences, Baku, Azerbaijan\\
$^{12}$ Institut de F{\'\i}sica d'Altes Energies and Departament de F{\'\i}sica de la Universitat Aut{\`o}noma de Barcelona and ICREA, Barcelona, Spain\\
% ...
$^{21}$ Physikalisches Institut, University of Bonn, Bonn, Germany\\
% ...
$^{48}$ Fakult{\"a}t f{\"u}r Mathematik und Physik, Albert-Ludwigs-Universit{\"a}t, Freiburg, Germany\\
$^{49}$ Section de Physique, Universit{\'e} de Gen{\`e}ve, Geneva, Switzerland\\
% ...
$^{88}$ Department of Physics and Astronomy, Michigan State University, East Lansing MI, United States of America\\
% ...
$^{105}$ Nikhef National Institute for Subatomic Physics and University of Amsterdam, Amsterdam, Netherlands\\
% ...
$^{111}$ Homer L. Dodge Department of Physics and Astronomy, University of Oklahoma, Norman OK, United States of America\\
$^{112}$ Department of Physics, Oklahoma State University, Stillwater OK, United States of America\\
% ...
$^{115}$ LAL, Universit{\'e} Paris-Sud and CNRS/IN2P3, Orsay, France\\
% ...
\end{flushleft}



\begin{thebibliography}{00}
\bibitem{coordinate}
The ATLAS reference system is a Cartesian right-handed coordinate system with its origin at the nominal interaction 
point~(IP) in the center of the detector and the $z$-axis along the beam direction. The $x$-axis points from the IP to the center of 
the LHC ring, and the $y$-axis points upward. Cylindrical coordinates $(r,\phi)$ are used in the transverse plane, $\phi$ being 
the azimuthal angle around the beam direction. The pseudorapidity is defined in terms of the polar angle $\theta$ as 
$\eta=-\ln\tan(\theta/2)$. Transverse momentum ($p_{\rm T}$) is defined relative to the beam axis.
\end{thebibliography}
\end{document}